\documentclass[12pt]{article}
\usepackage{amssymb,amsfonts,amsmath,amsthm,multirow,latexsym,lscape}
\usepackage[dvips]{graphicx}
\usepackage{pdflscape}
\usepackage{fullpage}
\usepackage{dcolumn}
\usepackage{natbib}
\usepackage{booktabs}
\usepackage{threeparttable}
\usepackage{authblk}
\usepackage{hyperref}
\usepackage{setspace}
\usepackage{indentfirst}
\usepackage{siunitx}
\usepackage[margin=1in]{geometry}
\usepackage{palatino}
\usepackage{xcolor}

\usepackage[hang, flushmargin]{footmisc}         
\usepackage{footnotebackref}

\newcommand{\tabnote}[2]{\begin{center}\parbox[c]{#1}{\footnotesize #2} \end{center}}

\newtheorem{assumption}{Assumption}

\makeatletter
\newcommand\org@hypertarget{}
\let\org@hypertarget\hypertarget
\renewcommand\hypertarget[2]{%
\Hy@raisedlink{\org@hypertarget{#1}{}}#2%
} \makeatother

\begin{document}

    \title{What can we learn about SARS-CoV-2 prevalence from testing and hospital data?%
    \thanks{We are grateful to Justin Blackburn, Seth Freedman, Aaron Kolb, Charles Manski, and  Kosali Simon for helpful comments on earlier drafts, to Evegenia Teal for assistance with the data, and to Agrayan Gupta for helpful research assistance. We acknowledge the Indiana University Pervasive Technology Institute for providing HPC resources that have contributed to the research results reported in this paper (\url{https://pti.iu.edu}). All errors are our own.}
}

\author{
Daniel W. Sacks\thanks{%
  Business Economics \& Public Policy, Kelley School of Business, Indiana University
  (dansacks@indiana.edu).}%
\and
Nir Menachemi\thanks{%
  Fairbanks School of Public Health and Kelley School of Business, Indiana University, and Regenstrief Institute, Inc.
  (nirmena@iu.edu).}%
\and
Peter Emb\'{i}\thanks{%
    School of Medicine, Indiana University, Indiana University Health System, and Regenstrief Institute, Inc.
  (pembi@regenstrief.org).}
\and
Coady Wing\thanks{%
  O'Neill School of Public and Environmental Affairs, Indiana University
  (cwing@indiana.edu).} %
}

\begin{spacing}{0.9}
\begin{titlepage}
\maketitle

\thispagestyle{empty}

\begin{abstract}
\noindent
Measuring the prevalence of active SARS-CoV-2 infections in the general population is difficult because tests are conducted on a small and non-random segment of the population. However, people admitted to the hospital for non-COVID reasons are tested at very high rates, even though they do not appear to be at elevated risk of infection. This sub-population may provide valuable evidence on prevalence in the general population. We estimate upper and lower bounds on the prevalence of the virus in the general population and the population of non-COVID hospital patients under weak assumptions on who gets tested, using Indiana data on hospital inpatient records linked to SARS-CoV-2 virological tests. The non-COVID hospital population is tested fifty times as often as the general population, yielding much tighter bounds on prevalence. We provide and test conditions under which this non-COVID hospitalization bound is valid for the general population.  The combination of clinical testing data and hospital records may contain much more information about the state of the epidemic than has been previously appreciated.  The bounds we calculate for Indiana could be constructed at relatively low cost in many other states.

\end{abstract}
\vspace{0.2in}

\end{titlepage}

\end{spacing}

     
    \section{Introduction} 
        \label{sec:introduction}
        Constructing credible estimates of the current prevalence of SARS-CoV-2 in the United States is challenging. Despite growing since the start of the epidemic, testing rates remain low in most of the country. Moreover, tests are often allocated to people exhibiting COVID-19 symptoms or who are thought to have come into contact with the virus \citep{wsjTest}. For example, New York and Texas both use a self-diagnostic tool to screen people for testing.%
\footnote{See \url{https://covid19screening.health.ny.gov/covid-19-screening/} and \url{https://txctt.force.com/ct/s/assessment?language=en_US}.} 
The low rate of testing in the general population means that the number of confirmed SARS-CoV-2 cases almost certainly understates the true number of infections in the population. At the same time, statistics like the fraction of tests that are positive likely overstate population prevalence because the tested population is more likely to be infected than the population as a whole. 

In this paper, we propose a new approach to measuring the point-in-time prevalence of active SARS-CoV-2 infections in the overall population using data on patients who are hospitalized for non-COVID reasons. There is less uncertainty about SARS-CoV-2 prevalence among non-COVID hospital patients because people in the hospital are tested at much higher rates than the general population, even if they are hospitalized for reasons unrelated to COVID-19 \citep{SuttonEtAl2020}. We show how a family of weak instrumental variable assumptions can be used to reduce the inferential uncertainty about the prevalence of SARS-CoV-2. We combine these assumptions with linked testing-hospital data from Indiana to estimate relatively tight upper and lower bounds on the prevalence of active SARS-CoV-2 infections in the overall population in Indiana in each week from mid-March to mid-December. The detailed Indiana data allows us to conduct robustness checks that partially validate some of our assumptions. Our basic method (without the validation checks) could be implemented using data that many states are already collecting and partially reporting. Thus, our approach could help states extract timely prevalence information using existing surveillance data. Importantly, our paper is focused on estimates of the fraction of the population that would test positive in each week. These estimates are distinct from recent efforts to estimate the share of the population \textit{ever infected} with SARS-CoV-2 \citep{ManskiMolinari2020}. The distinction between active prevalence and cumulative prevalence is important because the prevalence of active infections is a key determinant of the spread of the epidemic, given that the level of immunity in the population is thought to be quite low. 

Estimates and forecasts of the prevalence of active SARS-CoV-2 infections are crucial for public and private responses to the disease. They have shaped decisions about disruptive non-pharmaceutical interventions such as school closures, non-essential business closures, gathering restrictions, stay-at-home mandates, and temporary increases in the generosity of the unemployment insurance system \citep{gupta2020}. Reported estimates of prevalence likely also motivate individual precautionary behaviors ranging from wearing a mask to reducing demand for goods and services that require physical interaction \citep{gupta_brookings_2020,AllcottEtAl2020, philipson2000economic, philipson1996private, kremer1996integrating}. Finally, estimates of prevalence are a necessary input into efforts to measure other quantities of interest, like the infection fatality rate and the infection hospitalization rate.

Given its importance, researchers have developed several approaches to measuring SARS-CoV-2 prevalence given non-representative testing. One of the most credible is to conduct a biometric survey in which tests are offered to a representative sample of the population \citep{NirWave1, NirWave2,GudbjartssonEtAl2020}. Other studies have tested a census of smaller populations such as cruise ships  \citep{DiamondPrincessFeb2020}. However, it is difficult and costly to regularly implement a survey with accurate coverage and high response rates, especially a new survey that has not been in the field for long. A second approach involves backcalculation methods, which use data on observed hospitalizations or deaths to infer disease prevalence at earlier dates using assumptions about the unobserved parameters that determine the progression of the disease, hospitalization rates, and case fatality rates \citep{brookmeyer1988method, egan2015review, FlaxmanEtAl2020, SaljeEtAl2020}. Backcalculation may work well if hospitalizations or deaths are well measured, and if previous research has already reached consensus about key parameters related to the disease. However, backcalculation may be less credible for a novel virus because the scientific knowledge base is smaller. And even when backcalculation is  based on credible assumptions, it may be of limited value for public health decision making because both hospitalizations and deaths lag current infections \citep{VerityEtAl2020}.

An alternative to biometric surveys and backcalculation is to combine non-random clinical testing data with weak distributional assumptions to construct bounds on population prevalence (e.g. \cite{manski1999identification, wing2010three}). In the COVID-19 epidemic,  \citet{stock2020identification} provide bounds on population prevalence using a testing encouragement design, which is not always available. \citet{ManskiMolinari2020} bound the share of the population that has ever been infected under a ``test monotonicity assumption" that the infection rate is weakly higher among the tested than among the untested. This assumption is appealingly credible, and the bounds can be calculated from widely available test data. However, even when the focus is cumulative prevalence under test monotonicity, the prevalence bounds are often wide because testing is so rare. 

The strategy we pursue in this paper is based on the insight that test rates are especially high among hospitalized patients, even patients hospitalized  for reasons that are apparently unrelated to COVID-19, such as labor and delivery or vehicle accidents. The upper and lower bounds on prevalence in these populations will tend to be much tighter than similar bounds in the general population. In addition, it is plausible that some types of hospitalizations occur for reasons that are independent of infection risk. For example, people who are hospitalized for injuries sustained in a traffic accident might be expected to have about the same risk of SARS-CoV-2 infection as the general population. We build on this insight to estimate more informative upper and lower bounds on weekly SARS-CoV-2 prevalence in the population. 

Our paper makes methodological contributions that may be relevant to the development of more informative public health surveillance systems. We describe the conditions under which upper and lower bounds on active prevalence among non-COVID hospitalizations are valid estimates of the upper and lower bounds on prevalence in the general population. We maintain the test monotonicity assumption throughout, and we derive upper and lower bounds on prevalence in the population under two alternative assumptions about the representativeness of non-COVID hospitalizations for the broader population. The first assumption is a relatively weak ``hospital monotonicity" assumption that prevalence is at least as high among non-COVID hospital patients as it is in the general population. This assumption would be satisfied even if hospitalized patients are at greater risk of COVID because, for example, people who get into car accidents have more social interactions. The second assumption is a stronger "hospital independence" assumption that prevalence is the same among non-COVID hospital patients as it is in the general population. The resulting bounds are informative for prevalence at a point in time, not just cumulative prevalence. This is important because the information required to estimate the bounds is closely related to the simple statistics that many states already report. It appears possible for many states to use our method to report upper and lower bounds on prevalence in near real time using data that they already collect and report. In particular, states already report the COVID-hospitalization rate, as well as overall test rates and test positivity rates. To report a version of the upper and lower bounds we describe in this paper, states would simply have to compute testing and positivity rates among non-COVID hospitalizations.%
\footnote{Our methodological contribution is contemporaneous  work by \citet{gelman2021}. Like us, they study how non-COVID hospitalizations can improve COVID-19 surveillance. They study a hospital system which tests all of its patients and is therefore free of selection into testing. Whereas we develop bounds to account for selective testing, they focus on adjusting for demographic differences between hospitalized patients and the general community. In principle both approaches can be combined.}

Our first empirical contribution uses this framework to estimate  a collection of upper and lower bounds on weekly prevalence of SARS-CoV-2 in Indiana. To operationalize the basic idea, we work with two definitions of non-COVID hospitalizations. The first, which we call non-Influenza- or COVID-like-illness (non-ICLI) hospitalizations, simply excludes all hospitalizations with diagnosis for ICLI \citep{cdcCovidCodes,afhsc2015}. This definition would be easy to implement in many different hospital data sets and yields a large population of hospital patients.  However, we also work with a definition using a narrower set of patients who are hospitalized for six groups of clear non-COVID causes: (i) cancer; (ii) appendicitis and vehicle accidents; (iii) labor and delivery; (iv) AMI and stroke; (v) fractures, crushes, and open wounds; and (vi) other accidents. The clear-cause analysis is more intuitive and transparent, but it might be harder to implement as part of a public health surveillance system. In addition, the clear cause analysis often involves the analysis of smaller samples of data. In practice, this may create a trade-off between the inferential uncertainty that stems from the width of the bounds under strong vs weak assumptions, and the statistical uncertainty that arises when estimates are based on fewer observations. 

We show that, in Indiana, test rates are much higher among non-COVID hospital patients than in the general population. For example, in June, about 0.4 percent of the general population was tested in a given week, compared with 24 percent of non-COVID hospital patients. Test positivity rates are lower among non-COVID hospital patients than in the general population. In the general population, 4.5 percent of tests were positive. In contrast, among non-COVID hospital patients who were tested, only 2.2 percent of tests were positive. These testing and positivity rates can be combined to estimate upper and lower bounds on prevalence. Under the test monotonicity assumption, between 0.05 and 4.5 percent of the general population was infected with SARS-CoV-2 on June 15. Under the same test monotonicity assumption, the bounds are half as wide for non-COVID hospital patients: prevalence was between 0.7 and 2.2 percent.  

By the end of November, with the pandemic worsening, the test rate had increased to about 3 percent in the general population, and to more than 40 percent in the non-COVID hospitalization population. Although testing rates had increased in the population, the test positivity rate had also risen substantially. As a result, the test monotonicity bounds on prevalence in the general population widened to 0.7 to 22.5 percent in November. In contrast, among non-COVID hospitalization patients, the bounds at the end of November were 3.1 to 6.2 percent - unambiguously higher than in June. Low testing rates in the population mean that selection bias in testing creates so much ambiguity that it is impossible to confirm that that prevalence was rising using the state's testing data. In contrast, the bounds on prevalence based on non-COVID hospital patients are tight enough that they show an unambiguous rise in SARS-CoV-2 prevalence from summer to late fall. 

Under stronger assumptions, these tight hospital bounds can be interpreted as bounds on population SARS-CoV-2 prevalence. Specifically, under a hospital monotonicity assumption, the upper bound on prevalence in the non-COVID hospital population is a valid upper bound on population prevalence. And under the stronger hospital independence assumption, the upper and lower bounds on prevalence in the non-COVID hospital population are valid upper and lower bounds on population prevalence. 

In the results section of the paper we present bounds under alternative hospital representativeness assumptions (none, monotonicity, independence) to allow readers to make their own judgements about which assumptions are credible, and to better understand how much information about prevalence is derived from the data versus assumptions. To characterize the statistical uncertainty in our work, we estimated 95\% confidence interval around each set of upper and lower bounds using a version of the methods developed in the partial identification literature \citep{horowitz2000nonparametric,ImbensManski2004}.

We also report bounds based on the non-ICLI definition as well as multiple clear cause definitions and find that the two approaches give similar results. In addition, we estimated the upper and lower bounds on prevalence among ICLI hospitalizations. We find, of course, much higher prevalence among this group, but our bounds still rule out very high prevalence. Specifically, we find that SARS-CoV-2 prevalence among ICLI hospitalizations is no higher than 50 percent at its peak, and no higher than about 30 percent by mid-June. This shows that there is value to testing even highly symptomatic patients, as their SARS-CoV-2 rate is far from 100 percent, and testing outcomes would be informative for treatment and quarantine decisions. 

Our second empirical contribution is to assess the credibility of key assumptions that would be difficult to study in other data sets. \citet{ManskiMolinari2020} point out that the accuracy of SARS-CoV-2 virological tests is not well understood. Incorporating information about testing errors alters the bounds on prevalence. We use data on people who were tested twice in a two day period to shed some light on the fraction of people who test negative but are actually infected. We tentatively conclude that test errors have a negligible effect on the upper and lower bounds on prevalence reported in our paper.

We also assess the credibility of the hospital representativeness assumptions at the core of the paper. The most restrictive hospital independence condition assumes that SARS-CoV-2 prevalence is the same in the non-COVID and general populations, and the weaker hospital monotonicity condition assumes that prevalence is at least as high among the hospitalized. Although we are not able to directly validate these assumptions, we probe their credibility in three ways. First, we compute age-standardized estimates of the upper and lower bounds by estimating age group specific bounds and then averaging over the population age distribution. This helps address concerns that the hospital population may have a different age distribution than the general population. Second, we compare the hospitalization bounds to estimates of population prevalence obtained from tests of a random sample of Indiana residents in April and June \citep{NirWave1, NirWave2}. Third, we examine the pre-hospitalization test rates of non-COVID hospitalized patients. These validity checks are roughly consistent with the the hospital independence assumption, and highly consistent with hospital monotonicity. This suggests that these assumptions might be considered reasonable in other states where it would be easy to estimate the upper and lower bounds but harder to perform elaborate validation exercises.

Overall, our results indicate that combining testing data and information on non-COVID hospitalizations may be a feasible and informative way of measuring SARS-CoV-2 prevalence. In the most recent weeks of our data (and under test monotonicity but without hospitalization data), we can only conclude that at most 20 percent of the overall Indiana population was actively infected. With hospitalization data and the hospital monotonicity assumption, we can conclude that at most 256 percent of the Indiana population was infected.  The bounds are tight enough that we can conclude that prevalence rose unambiguously from mid-summer to late fall. Similar bounds could be constructed in other states using aggregate data on non-COVID hospitalizations and their testing and test positivity rates, potentially improving SARS-CoV-2 surveillance systems across the country.

    \section{Inferring COVID Prevalence from Incomplete Testing} 
        \label{sec:identification}
        
The empirical goal of our study is to establish upper and lower bounds on the prevalence of active SARS-CoV-2 infections in the Indiana population in each week. Measuring prevalence is challenging because only a small fraction of people are tested  in any given week. One way to overcome this challenge is to combine the data with assumptions about testing and infection risk that are strong enough to point identify prevalence. For example, under the assumption that the people who are tested are a representative sample from the overall population, prevalence among the tested equals the population prevalence. However this assumption is not very credible because of non-random selection into testing, which may occur because people who exhibit symptoms are tested at higher rates. In this paper, we explore the identifying power of weak instrumental variable assumptions related to hospital based testing. These assumptions are reasonably credible, but they only partially identify prevalence. This means that the assumptions establish upper and lower bounds on prevalence but do not restrict it to a single value. 

The instrumental variable methods we use are standard in the literature on partial identification econometrics \citep{manski1999identification, manski2009identification}. And they are closely related to methods from research on sample selection bias (e.g. \citet{heckman1979sample}). But the approach may seem unfamiliar to readers who mainly work in the context of causal inference  (e.g. \citet{AngristPischke2008}). In the causal inference literature, an instrumental variable is a factor which affects treatment exposure but is mean-independent of potential outcomes and satisfies an exclusion restriction. In the sample selection literature, an instrumental variable affects sample selection, but is mean-independent of the outcome of interest and satisfies an exclusion restriction. The classic application involves the distribution of wages among women. Wages are observed for women who work, and unobserved for women who do not work. An instrumental variable is something that affects a woman's employment, but is not associated with her wage offer. Our application is conceptually similar: SARS-CoV-2 status is observed for people who are tested, and unobserved for people who are not tested. An instrument would be a variable that affects testing but is not otherwise associated with SARS-CoV-2 infection risk. The classical sample selection literature combines instrumental variable assumptions with a parametric model of the selection decision and the outcome to point identify parameters of the outcome distribution. The partial identification literature is non-parametric and relaxes the functional form assumptions. It also examines weaker forms of instrumental variable assumptions. For example, in this paper we consider replacing mean independence assumptions weak sign restrictions, generating ``monotone" rather than mean independent instrumental variables  \citep{manski2000monotone, Manski2020}. All of these different approaches are ``instrumental variable methods" because they rely on assumptions to restrict the relationship between an instrument and an outcome. 

\begin{figure}[h!]

    \caption{Stronger assumptions generate tighter prevalence bounds }
    \label{fig:flowchart}
    \begin{center}
    \centering \includegraphics[width=.85\textwidth]{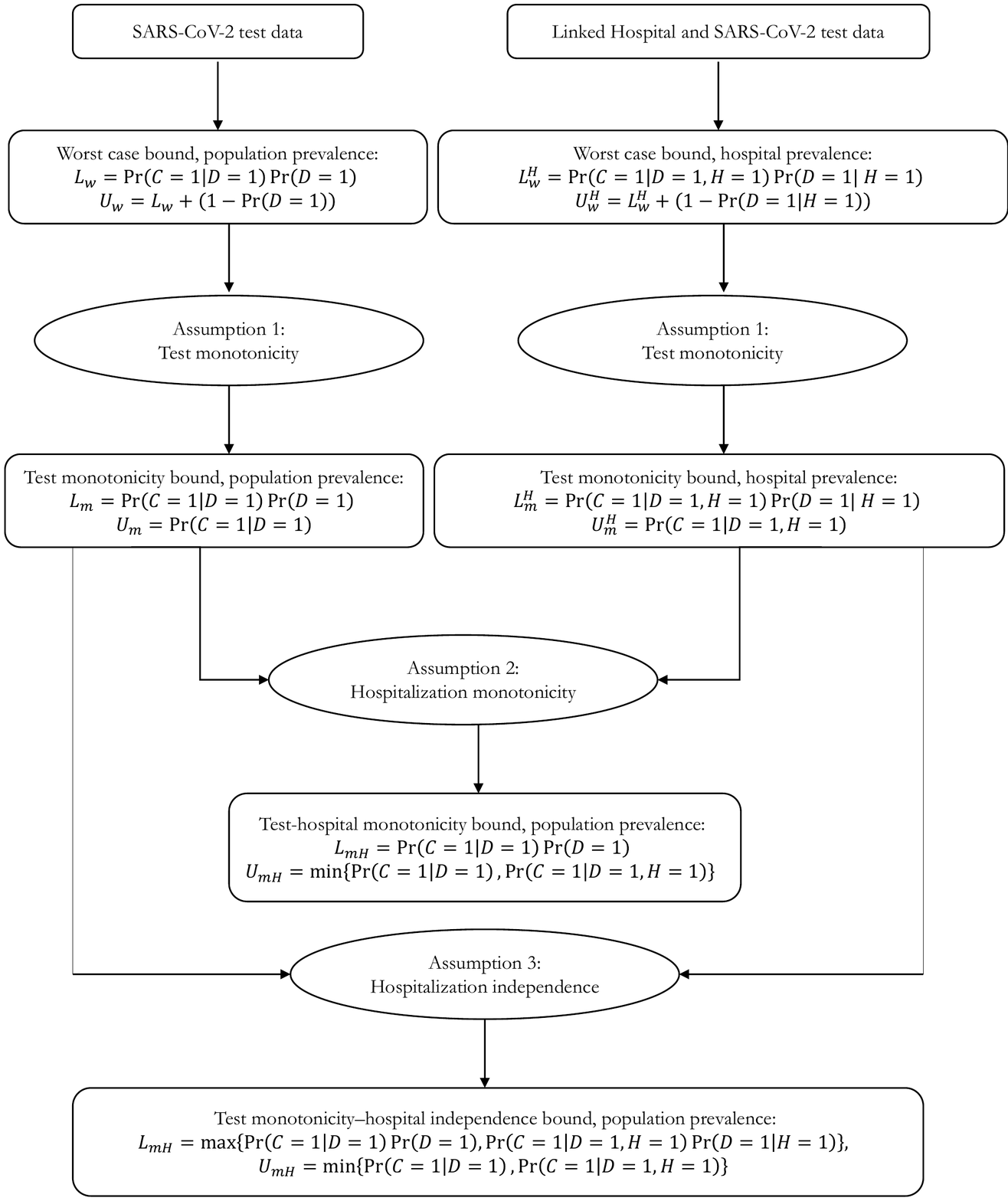}
    \end{center}
    \tabnote{\textwidth}{Notes: Figure illustrates how increasingly strong assumptions generate successively tighter bounds on prevalence.  Without any assumptions the data yield the worst case bounds. With a test monotonicity assumption, we bound population prevalence and hospital prevalence. With hospitalization monotonicity, the hospital bounds are informative for population prevalence. With hospital independence, the bounds tighten further. The flowchart maintains the assumption of no measurement error throughout, for reasons described in Section \ref{test_error}.}
\end{figure}

An appealing feature of the partial identification approach is that inferences can be tightly linked to assumptions, making it transparent how stronger assumptions can generate tighter inferences. Our approach introduces two monotone instrumental variable assumptions: test monotonicity and hospitalization monotonicity. We show how to use these relatively weak assumptions to narrow the worst-case (no-assumption) bounds on prevalence. To tighten the bounds further, we introduce a stronger assumption of hospital independence. \hyperref[fig:flowchart]{Figure \ref*{fig:flowchart}} gives a schematic representation of the data, assumptions, and results we present in the study. The figure shows how our assumptions and data are combined to generate inferences on SARS-CoV-2 prevalence, with increasingly strong assumptions yielding increasingly tight bounds. After introducing, discussing, and justifying our key assumptions, we return to the figure to summarize how different assumptions yield different bounds. 

\subsection{Notation and Worst Case Bounds}

We use $i=1...N$ to index members of the population of Indiana. Let $C_{it} = 1$ indicate that person $i$ is currently infected with SARS-CoV-2 on date $t$. Leaving conditioning on the date implicit to reduce clutter, the population prevalence of active SARS-CoV-2 infections in Indiana at date $t$ is $Pr(C_{it} = 1) = \frac{1}{N} \sum_{i=1}^{N}C_{it} $. We are also interested in prevalence among hospital inpatients with various COVID- and non-COVID-related diagnoses.  Let $H_{it}$ be a binary indicator set to 1 if the person was hospitalized with a specified non-COVID-related diagnosis. Then $Pr(C_{it} = 1 | H_{it} = 1) $ is the prevalence of active SARS-CoV-2 infections in the sub-population of people who were admitted with a non-COVID-related diagnosis on date $t$. 

The central problem in estimating prevalence is that values of $C_{it} $ are unknown for most people on most days, because testing is rare. Let $D_{it} = 1$ if person $i$ was tested on $t$ and $D_{it} = 0$ if the person was not tested. $Pr(D_{it} = 1) $ is the proportion of the population tested on date $t$, where conditioning on $t$ is implicit. Continuing with the notation laid out above, $Pr(C_{it} | D_{it} = 1)$ and $Pr(C_{it} | D_{it} = 0)$ represent the prevalence of SARS-CoV-2 among people who are tested and not tested, respectively. The value of $C_{it} $ is observed for people with $D_{it} = 1 $, but unknown for people with $D_{it} = 0$. This means that $Pr(C_{it} | D_{it} = 0)$  is not identified by the data on testing and test outcomes. Uncertainty about the prevalence of SARS-CoV-2 among people who have not been tested is the main reason why it is difficult to estimate population prevalence using data from clinical tests. 

In the absence of any distributional assumptions, the observed clinical tests partially identify prevalence overall, and in any sub-populations that can be defined by observable covariates. To see the point, use the law of total probability to decompose population prevalence: 

\begin{equation}
\label{eq:tp}
    Pr(C_{it}=1)  = Pr(C_{it} =1| D_{it}=1)  Pr(D_{it}=1) 
    +Pr(C_{it}=1 | D_{it}=0) Pr(D_{it}=0)
\end{equation}

\noindent The only unknown quantity on the right-hand side of the expression is  $ Pr(C_{it}=1|D_{it} = 0)$, which is prevalence among people who were not tested. All that is known is that this value lies between $0$ and $1$. Substituting 0 and 1 for the unknown prevalence yields worst-case lower and upper bounds $L_w$ and $U_w$ on population prevalence:

\begin{alignat*}{2}
L_{w} & =  \underbrace{Pr(C_{it}=1|D_{it} = 1) Pr(D_{it} = 1)}_\text{Confirmed Positive Rate} \\
U_{w} &=   \underbrace{Pr(C_{it}=1|D_{it} = 1) Pr(D_{it} = 1)}_\text{Confirmed Positive Rate} 
+ \underbrace{Pr(D_{it} = 0)}_\text{Untested Rate} .
\end{alignat*}

\noindent These bounds defines the set of values for prevalence that are compatible with the observed data. From the definition of joint and conditional probability, we can rewrite the lower bound as the joint probability that a person is both tested and infected: $L_{w} = Pr(C_{it} = 1, D_{it})$. In other words, the worst case lower bound is the fraction of the population that has actually tested positive for SARS-CoV-2 on a given date. Since this group of people is definitely infected, the lower bound is the \textit{confirmed positive rate} and it makes intuitive sense that overall prevalence cannot be any lower than confirmed positive rate. The lower bound is the population prevalence in a scenario where none of the untested people are infected. The upper bound gives the opposite extreme in which all of the untested people are infected. Put differently, the worst case upper bound is the \textit{confirmed positive rate plus the untested rate}. The components of the bounds can be estimated using the appropriate proportions in test and hospital data. To account for statistical uncertainty in the estimates of the upper and lower bounds, we use methods developed in \citet{horowitz2000nonparametric} and \citet{ImbensManski2004} to construct confidence intervals on the bounds. See \hyperref[app:ci]{Appendix \ref{app:ci}} for details. 

\subsection{Test monotonicity} 
\label{sec:test_monotonicity}

The worst case bounds are very wide when testing rates are low. To narrow the bounds, \cite{ManskiMolinari2020} propose the ``test monotonicity" condition. In our context, test monotonicity requires 

\begin{assumption}
\label{test_monotonicity_assumption}
(Test monotonicity) $Pr(C_{it}=1|D_{it}=1) \geq Pr(C_{it}=1|D_{it} = 0)$
\end{assumption}

\noindent Assumption \ref{test_monotonicity_assumption} requires that the prevalence of SARS-CoV-2 is at least as high in the tested population as it is in the untested population. This assumption is untestable because prevalence is unknown in the tested sub-population. However, the assumption has credibility because virological tests are typically allocated to symptomatic individuals, who have a higher than average likelihood of infection. To apply the assumption, recall that the identification problem is that prevalence in the untested sub-population -- $Pr(C_{it}=1 | D_{it}=0)$ -- is unknown. Assumption \ref{test_monotonicity_assumption} implies that $0 \leq Pr(C_{it} =1| D_{it}=0) \leq Pr(C_{it=1} | D_{it}=1)$. Substituting the values $0$ and $Pr(C_{it} =1| D_{it}=1)$ into \hyperref[eq:tp]{Equation \ref*{eq:tp}} yields a new set of bounds on overall prevalence:
\begin{align*}
L_{m} & =  \underbrace{Pr(C_{it}=1|D_{it} = 1) Pr(D_{it} = 1)}_\text{Confirmed Positive Rate},\\
U_{m} & = \underbrace{Pr(C_{it}=1|D_{it} = 1)}_\text{Test Positivity Rate}.
\end{align*}
\noindent The new upper bound is the prevalence in the tested population, which is often called the \textit{test positivity rate}. This new test monotonicity upper bound will be lower than the worst-case upper bound as long as prevalence in the tested sub-population is less than 1.  In our data, test rates are often less than 1 percent and positivity rates in the population are often  10 percent or less, so this assumption brings the upper bound down from 99 percent to 10 percent or less. A similar assumption could be made for the non-COVID hospitalization population, yielding bounds $L_{m}^H$ and $U_{m}^H$ on prevalence among the non-COVID hospitalized population. 

Test monotonicity appears to be a credible assumption given the reliance on symptomatic testing for SARS-CoV-2 in the United States. Nevertheless, it is worth considering scenarios under which the assumption would fail. In the abstract, test monotonicity could fail if the demand for voluntary testing is higher among people with lower risk of infection. The ``worried well'' phenomenon that occurred the HIV/AIDS epidemic is one example \citep{cochran1989women}. Theoretically, a worried well effect could arise because of preferences that increase a person's demand for testing are correlated with preferences that reduce the person's SARS-CoV-2 infection risks. A worried well could also arise if testing is cheaper or more convenient for sub-populations that are actually at low risk of infection. This could happen -- for example -- because of socio-economic, racial, or geographic disparities in the severity of the epidemic and available testing capacity. Although these scenarios are plausible, they seem unlikely to dominate the effects of symptomatic testing procedures that support test monotonicity.

\subsection{Inferring Population Prevalence From Non-Covid Hospital Patients}
\label{sec:monotone_hosp}

Test monotonicity can be used to narrow the the bounds on prevalence in the population and also among non-COVID hospitalized patients. Because testing rates are much higher in hospitals than in the general population, the bounds on prevalence in hospitalized sub-populations are much narrower. Thus, assumptions that link hospital and population prevalence may be a powerful way to reduce uncertainty about population prevalence. We pursue two types of assumptions that enable extrapolation from non-COVID hospital populations to the general population: (i) hospitalization monotonicity and (ii) hospitalization independence. These are both forms of \textit{hospital instrumental variable} assumptions, and we refer to them collectively as hospital IV assumptions.

\subsubsection{Hospitalization Monotonicity}
\label{sec:monotone_hosp}

In some contexts, it may be plausible to assume that people who are hospitalized for a non-COVID related health condition will not have a lower risk of SARS-CoV-2 infection than than the general population. Stated formally, the hospitalization monotonicity assumption is:

\bigskip

\begin{assumption} (Hospitalization Monotonicity)
\label{hospital_monotonicity_assumption}
$Pr(C_{it}=1 | H_{it}=1) \geq Pr(C_{it}=1)$
\end{assumption}

\bigskip

\noindent Assumption \ref{hospital_monotonicity_assumption} requires that the prevalence of active SARS-CoV-2 infections among non-COVID hospital patients is at least as high as the prevalence of active infections in the general (non-hospitalized) population. When prevalence is bounded in the hospitalized and general populations, the hospital monotonicity assumption may further reduce the width of both sets of bounds by ruling out values that would violate the restriction. To see the idea, suppose that we apply Assumption \ref{test_monotonicity_assumption} in both the hospitalized population and the general population. Let $U_{m}^{H}$ and $L_{m}^{H}$ represent the upper and lower bounds on SARS-CoV-2 prevalence in the hospitalized sub-population under the test monotonicity assumption. Following the notation above, $U_m$ and $L_m$ represent the test monotonicity bounds in the general population. Layering the additional hospitalization monotonicity assumption, (Assumption \ref{hospital_monotonicity_assumption}) creates a cross-population restriction that implies that the upper bound on \textit{population prevalence} $(U_m)$ cannot be larger than the upper bound on \textit{hospital prevalence} $(U_{m}^{H})$. Maintaining both Assumption \ref{test_monotonicity_assumption} (test monotonicity) and Assumption \ref{hospital_monotonicity_assumption} (hospitalization monotonicity) yields the following upper bound on SARS-CoV-2 prevalence in the population:
\begin{align*}
U_{m,h} & = \min \left\lbrace U_{m}, U_{m}^H \right\rbrace  = \min\left\lbrace Pr(C_{it}=1|D_{it}=1), Pr(C_{it}=1|D_{it}=1, H_{it}=1)\right\rbrace \\
&= \min\left\lbrace \text{Population test positivity}, \text{Hospital test positivity}\right\rbrace ,
\end{align*}
\noindent In our data, the hospital upper bound is typically lower than the population upper bound, so the hospital monotonicity condition implies that the positivity rate among non-COVID hospitalizations is an upper bound population prevalence. In principle, the monotone hospitalization assumption could be used to derive a potentially higher lower bound on prevalence in the hospital population. But this lower bound condition is essentially non-binding in practice, so we ignore it here.

\subsubsection{Hospitalization Independence}
\label{sec:hosp_ind_section}

In some situations it may be credible to assume a stronger condition than monotone hospitalizations, that hospitalization for a non-COVID health condition is mean-independent of SARS-CoV-2 infection risk. Formally, the hospitalization independence assumption can be written:

\bigskip

\begin{assumption} (Hospitalization Independence)
\label{hospital_independence_assumption}
$Pr(C_{it} =1| H_{it} = 1) = Pr(C_{it}=1)$
\end{assumption}

\bigskip
\noindent 

The independence assumption is stronger than the weak sign restriction embodied by the hospitalization monotonicity assumption. Assumption \ref{hospital_independence_assumption} implies that people who are hospitalized for a specified non-COVID health condition have the same probability of being infected with the virus as the general population. An equivalent statement of Assumption \ref{hospital_independence_assumption} is that people who are infected with SARS-CoV-2 have the same probability of being hospitalized for a non-COVID condition as people who are not infected with SARS-CoV-2, which implies  $Pr(H_{it} = 1 | C_{it}=1) = Pr(H_{it} = 1 |C_{it}=0) $. 

Under the hospital independence assumption, the bounds on population prevalence are defined by the intersection of the hospital and population bounds. As before, $U_m$ and $L_m$ are the upper and lower bounds on prevalence in the general population under the test monotonicity assumption alone. Likewise, $U_{m}^{H}$ and $L_{m}^{H}$ are the upper and lower bounds on prevalence in the hospital population under test monotonicity. Assumption \ref{hospital_independence_assumption} implies that prevalence is the same in the hospitalized and general populations. That implies that the lower bound on prevalence in the general population must be no lower than the larger of the two lower bounds. By the same logic, the upper bound on prevalence in the general population must be no larger than the smaller of the two upper bounds. Formally, layering the hospital independence assumption on top of the test monotonicity assumption yields a new set of lower and upper bounds:
\begin{align*}
L_{m,ind} & = \max \left\lbrace L_{m}, L_{m}^H \right\rbrace  \\
 &= \max \left\lbrace Pr(C_{it}=1|D_{it}=1)Pr(D_{it}), Pr(C_{it}=1|D_{it}=1, H_{it}=1)Pr(D_{it}) \right\rbrace, \\
U_{m,ind} & = \min \left\lbrace U_{m}, U_{m}^H \right\rbrace  \\
 &= \min\left\lbrace Pr(C_{it}=1|D_{it}=1), Pr(C_{it}=1|D_{it}=1, H_{it}=1)\right\rbrace.
\end{align*}
\noindent As it turns out, $U_{m,ind} = U_{m,h}$, so that the upper bound on prevalence is the same under \ref{test_monotonicity_assumption} and \ref{hospital_independence_assumption} as it is under Assumption \ref{test_monotonicity_assumption} and Assumption \ref{hospital_monotonicity_assumption}. What the hospital independence assumption buys us is a potentially tighter lower bound, which is now the greater of the lower bounds on population and hospital prevalence under test monotonicity. In practice we find that the lower bound is always higher in the non-COVID hospitalization sub-population than in the general population, so in practice this assumption implies that the lower bound on population prevalence is the confirmed positive rate among non-COVID hospitalizations.

\subsubsection{Hospital Instrumental Variables In Perspective}
\label{sec:hosp_iv_perspective}
Assumptions \ref{hospital_monotonicity_assumption} and \ref{hospital_independence_assumption} are strong instrumental variable restrictions that may hold for some non-COVID hospitalizations and not others. The independence assumption, Assumption \ref{hospital_independence_assumption}, is the strongest condition. It would be satisfied if non-COVID hospitalizations occur at random in the population, from the perspective of SARS-CoV-2 infection risk. For example, people who are hospitalized for cancers that are mainly due to genetic rather than behavioral factors are unlikely to be selected on differential SARS-CoV-2 risk. And people who are hospitalized because of car accidents are likely a fairly representative draw from the population of regular drivers. Hospitalization for stroke and heart attack also seems plausibly unrelated to a person's SARS-CoV-2 risk exposure. People who are in the hospital to deliver a baby or who are having ``routine" inpatient procedures like joint replacement are selected on the basis of decisions and behaviors that are mainly determined before the epidemic. They too might be expected to have the same distribution of SARS-CoV-2 infection risks as the general population.

Although hospital independence is plausible for some health conditions, it is also easy to think of reasons why it might fail. The basic demographics of people with particular health conditions may differ from the general population. This is easiest to notice for specific health conditions. Young people will be underrepresented among stroke and heart attack patients. Pregnant women are selected on gender. And older women will be underrepresented among pregnant women. In the empirical section of the paper, we report estimates of age-standardized upper and lower bounds that correct for differences in the age distribution across different hospital samples, which addresses some of these issues. But a more general concern with the independence assumption is that people who are hospitalized are simply sicker and more exposed to a variety of health risks than the general population. Furthermore, some types of hospital events may occur in part because people are engaged in activities that are higher risk during the epidemic. For example, people who drive regularly might have higher SARS-CoV-2 risk than the general population because they have more social interactions. 

These concerns motivate our effort to move away from full independence and consider the weaker hospitalization monotonicity condition, Assumption \ref{hospital_monotonicity_assumption} (monotone hospitalization). Of course, it is possible that even this weaker condition does not hold for some types of non-COVID hospitalizations. For instance, it would fail for non-COVID hospitalizations that are generated by a health condition that led a person to be more cautious about avoiding SARS-CoV-2 infection than the general population. In that case, we would expect the hospitalized group to have lower infection rates than the general population. While our discussion has focused on how individual behaviors might cause violations of our identifying assumptions, we emphasize that  Assumptions \ref{test_monotonicity_assumption}, \ref{hospital_monotonicity_assumption}, and \ref{hospital_independence_assumption} are restrictions on population and sub-population averages. The existence of a single person whose behavior or characteristics are counter to the assumption does not necessarily imply a violation of the population level condition. Our approach is to present upper and lower bounds under a range of weak and strong assumptions and for multiple non-COVID hospital sub-populations. Readers can inspect the evidence and judge for themselves which assumptions are most plausible and what that implies about the prevalence of COVID-19 during the epidemic. 

\subsection{Measurement Error in Testing}
\label{test_error}

Virological tests for the presence of SARS-CoV-2 may not be perfectly accurate, and so far there are no detailed studies of the performance of the PCR tests that Indiana is using to test people for SARS-CoV-2. To clarify how error-ridden tests complicate our prevalence estimates, we augment the notation to distinguish between test results and virological status. We continue to use $C_{it}$ and $D_it$ to represent a person's true infection and testing status at date $t$. But now we introduce $R_{it}$, which is a binary measure set to 1 if the person tests positive and 0 if the person tests negative. Using this notation, $Pr(C_{it} = 1 | D_{it} = 1, R_{it} = 1)$ is called the Positive Predictive Value (PPV) of the test among people who are tested and who test positive. $Pr(C_{it} = 0 | D_{it} = 1, R_{it} = 0)$ is called the Negative Predictive Value (NPV) among people who are tested and who test negative. $1-NPV = Pr(C_{it} = 1 | D_{it} = 1, R_{it} = 0)$ is the fraction of people who test negative who are actually infected with SARS-CoV-2.

Our initial worst case bounds assumed no test errors. Relaxing that assumption yields a different set of upper and lower bounds on prevalence. Following \citet{ManskiMolinari2020}, we assume that (i) $PPV = 1$ so that none of the positive tests are false, but (ii) $Pr(C_{it} = 1 | D_{it} = 1, R_{it} = 0) \in [\lambda_{l},  \lambda_{u}]$. The second condition imposes a bound on  $1-NPV$, which is the fraction of people who test negative who are actually infected. Under these two restrictions, the new worst case bounds work out to:
\begin{align*}
    L_{w,\lambda} &= L_{w} + \lambda_{l}Pr(R_{it} = 0 | D_{it} = 1)Pr(D_{it} = 1) \\
    U_{w,\lambda} &= U_{w} + \lambda_{u}Pr(R_{it} = 0 | D_{it} = 1)Pr(D_{it} = 1)
\end{align*}

Allowing for test errors increases the worst case lower bound by the best-case fraction of missing positives, and increases the worst case upper bound by the worst-case fraction of missing positives. Similar expressions hold for prevalence bounds under test monotonicity and other independence assumptions. 

The upshot is that knowledge of test accuracy is important for efforts to learn about prevalence. In their study of the cumulative prevalence of SARS-CoV-2 infections, \citet{ManskiMolinari2020} computed upper and lower bounds on prevalence under the assumption that $\lambda_{l} = .1$ and $\lambda_{u}=.4$, citing \cite{peci2014performance}. \citet{ManskiMolinari2020} view this choice of $.1 \leq 1-NPV \leq .4$ as  an expression of scientific uncertainty about test errors,  and they refer to the resulting prevalence bounds as ``illustrative.'' However, the structure of the test error bounds makes it clear that assumptions about the numerical magnitude of test errors have inferential consequences. For example, setting $ \lambda_{u}=.4$ implies that, regardless of the outcome of the test, at least 40 percent of the people who are tested for SARS-CoV-2 are infected.

Although there is little published evidence on the properties of the SARS-CoV-2 PCR test, previous research suggests that PCR test errors are uncommon in other settings. For example, \citet{peci2014performance} study the performance of rapid influenza tests using PCR-based tests as a \textit{gold standard}. PCR tests are used as a gold standard because they are expected to have very high PPV and NPV.

To shed more light on test errors, we constructed a sample of people who are tested and retested in a short interval, specifically people who were (i) tested on day $t$, (ii) not tested on day $t-1$, and (iii) were tested again on day $t+1$. We show in \hyperref[app:npv]{Appendix \ref*{app:npv}} how these data can be used to estimate error rates, under assumptions of random retesting and no false positives. Our data include 835,000 test-retest events. Using $R1_i$ and $R2_i$ to represent the results of a person's first and second test, we found that $Pr(R1_i = 1, R2_i = 1) = .11 $ and $Pr(R1_i = 0, R2_i = 0) = .88$ among the people in the twice-tested sample. The two tests were discordant for less than 1 percent of the twice-tested sample. These results imply a negative predictive value of 99.8 percent. 

This estimate of NPV depends on our assumptions of random retesting and no false positives. While the no false positive assumption appears plausible, random retesting is not necessarily satisfied. In particular, a patient with a suspected COVID case who initially tests negative may be retested; this selective retesting would bias us towards finding false negatives. Another reason for retesting is delays in processing results. If a patient was tested prior to a planned hospitalization, and the result is not available at the time of the hospitalization, the attending physician may order an in-hospital test, which would be available within hours. This type of retesting is less likely to lead to bias. As we explain in \hyperref[app:npv]{Appendix \ref*{app:npv}}, we can test for selection into retesting by looking for symmetry in test results. Under random retesting (and no false positives), the sequences ``positive-then-negative" and ``negative-then-positive" should be equally likely. In practice we find that ``negative-then-positive" is slightly more common, meaning that our test-retest sample likely disproportionately selects people with initial false negatives.

Overall, we think that a plausible value for $\lambda_{l}$ is nearly zero, and a plausible value for $\lambda_{u}$ is 0.005. Accounting for test errors in this range would have almost no effect on the upper and lower bounds reported in the paper. Test-retest data are potentially informative about test errors, but a limitation of is that retested people are not necessarily representative of the population. 
\subsection{Summary and data requirements} 

The analysis above shows how to combine test and hospital data with increasingly strong assumptions to obtain bounds on population prevalence. The overall approach requires data on the tested population that can be linked to hospital inpatient records that ideally include diagnosis codes to identify different types of hospitalizations. The diagnosis information is important because the hospitalization instrumental variable assumptions are more plausible for some types of hospitalizations. At a minimum, it is important to exclude non-COVID hospitalizations.

\hyperref[fig:flowchart]{Figure \ref*{fig:flowchart}} summarizes our methodological results and serves as a guide for interpreting our empirical findings. We work with three main assumptions: two weak monotonicity assumptions, and one conditional independence assumption. The flow chart shows which assumptions yield which bounds on population prevalence. Using only data and no assumptions, we have  worst-case bounds for prevalence in the general population and for hospitalized sub-populations. Under Assumption \ref{test_monotonicity_assumption}, test monotonicity, we have tighter bounds on prevalence for both populations. 

Assumptions \ref{hospital_monotonicity_assumption} and \ref{hospital_independence_assumption} let us extrapolate from the hospitalized sub-population to the general population. Under Assumption \ref{hospital_monotonicity_assumption}, hospitalization monotonicity, the upper bound on population prevalence tightens to the least upper bound under test monotonicity among the hospitalized subpopulation and the general population. Under Assumption \ref{hospital_independence_assumption}, hospitalization independence, the lower bound on population prevalence tightens to the greatest lower bound among the hospitalized subpopulation and the general population.

The bounds turn out to be fairly simple objects. Under test monotonicity the lower bound is the confirmed positive rate, the share of the population that tests positive. The upper bound under monotonicity is the test positivity rate, the share of tests that are positive. Under hospital monotonicity, the upper bound in the general population becomes the test positivity rate among non-COVID hospitalizations. And under hospital representativeness, the lower bound in the general population becomes the confirmed positive rate among the non-COVID hospitalized subpopulation. When presenting our results, we show confirmed positive rate and test positivity rates for the overall population and for the non-COVID hospitalized subpopulation. Thus readers can easily calculate bounds under any assumption, such as  test monotonicity alone or combined with hospitalization monotonicity.  

An appealing feature of these bounds is that they can be calculated with little additional data beyond what public health organizations already report. Every state already reports the number of tests and the number of positive tests, and many states report the number of COVID-related hospitalizations.%
\footnote{See, e.g., \cite{covidTracking}.} %
States would only have to report test and positivity rates for non-COVID-related hospitalizations. This appears possible because many states already report "suspected" or "under investigation" COVID hospitalizations, defined as hospitalized patients exhibiting COVID-like illness. Some states actually report both the number of hospitalizations of patients with COVID- or influenza-like illness and, separately, the number of hospitalizations of patients with a positive SARS-CoV-2 test (e.g. Arizona and Illinois \citep{azDashboard,ilHosp, ilSyndromic}).%
\footnote{States reporting both confirmed SARS-CoV-2 hospitalizations and hospitalizations of suspected cases or cases under investigation include California \citep{caDashboard}, Colorado \citep{coDashboard}, Mississippi \citep{msDashboard}, Tennessee \citep{tnDashboard}, and Vermont \citep{vtDashboard}. } 
Thus states have the capacity to identify ICLI-related hospitalizations and link hospitalization and testing data.

    \section{Indiana Hospital and Testing Data} 
        \label{sec:data}
        
\subsection{Data sets}
\label{sec:data_sets}
Our analysis is based on two main data sources managed by the Regenstrief Institute. First, we obtained data on the near universe of polymerase chain reaction (PCR)  tests for SARS-CoV-2 conducted in Indiana between January 1, 2020 and December 18, 2020. Second, we obtained data on all inpatient hospital admissions from hospitals that belong to the Indiana Network for Patient Care (INPC), which is a health information exchange that centralizes and stores data from health providers across the state of Indiana, including all hospitals with emergency departments.%
\footnote{See \cite{grannisINPC2005} for more details.} The hospital data are derived from the same database that the state uses for reporting hospitalizations on its dashboard \citep{inCovidDashboard}. We linked the individual testing and hospital inpatient records using an encrypted common identifier. Of course, only a subset of hospital patients are tested and only a subset of tested people appear in the inpatient hospital data. For both data sets, we are also able to link the individual testing and inpatient records with basic demographic data collected by the INPC; this information is available only for a subset of patients.

The test data contain individual records for nearly all of the SARS-CoV-2 tests conducted in Indiana during 2020. A small number of tests are excluded from our data because some institutions that conduct tests provide data to INPC but do not allow the data to be used for research purposes, and some tests are not included in our sample because the data are reported with a delay. The consequence of these exclusions is that we are missing some tests, which will result in a reduced lower bound in our framework. Despite these exclusions, our data set tracks the state's official case counts quite closely until November, 2020, when the state count exceeds ours; see \hyperref[fig:compare_to_state]{Appendix Figure \ref*{fig:compare_to_state}}. As an aggregate summary, our data contain 383,976 people with a positive test as of December 15, 2020, whereas the state reports 439,916  \citep{mphData}.%
\footnote{We compare positives rather than tests because some testing institutions appear not to report negative tests to the state. (Such institutions are not included in our data.)}
Each test record in our testing data includes information on the date the test specimen was obtained, the outcome of the test (positive, negative, or inconclusive), and a patient identifier that we use to link the test data to demographic files and inpatient hospital files. 

The hospital inpatient data contain separate observations for each admission. We always observe admission time and a patient identifier that we use to link to the test data and inpatient files. Discharge time and diagnosis information are only observed for a subset of admissions. (Not all fields are available for all admissions because different institutions contribute different information to the INPC.) Because the INPC data come from health care providers and payers, the same hospitalization can appear in the data set multiple times. To de-duplicate these records, we keep one observation per admission time (defined second-by-second), keeping the observation with the most diagnosis codes. This procedure could in principle result in us dropping relevant diagnostic information, but in practice this happens only very rarely. There are 707,734 admissions that are unique in terms of patient identifier and time, and 613,036 non-unique admission records. In 788 of these non-unique cases are there multiple records with the same number of diagnostic codes; we keep one of these multiples at random. In 122 of the 613,064 cases there are multiple records with diagnosis codes that provide conflicting first diagnoses.

\subsection{Measuring tests and cases}

\textbf{In-hospital testing, positivity rate, and confirmed positives} The fraction of people who are tested in the hospital is an important quantity of interest in our analysis because hospital patients are tested at a higher rate than the general population, and because hospital testing may be less correlated with COVID symptoms. Our data do not distinguish whether a person was tested in the hospital or whether the test was initiated independently of the hospital visit, but we can match tests to hospitalizations based on the test date and hospitalization date. We say that a hospitalized patient is tested in-hospital if she had at least one SARS-CoV-2 test dated between 5 days prior to admission to 1 day after admission, and we say she had a positive if she had at least one positive test in that window. Several considerations justify this definition. First, we want to include tests that are part of the admission. For patients with planned procedures, these tests may happen a few days prior to admission. For patients admitted from the emergency department, these tests may happen on the day of admission or the day after. Consistent with this view,  \hyperref[fig:hospital_test_times]{Appendix Figure \ref*{fig:hospital_test_times}}, shows that among non-ICLI hospitalizations, test rates begin to rise several days before admission and fall rapidly 3-4 days after admission. Second, we want keep the post-admission window relatively short to ensure that we do not pick up hospital-acquired SARS-CoV-2 infections, although this may not be a major concern since \cite{rhee2020incidence} indicate that hospital-acquired SARS-CoV-2 infections are quite rare. Finally,  we limit the window to seven days, so that we can compare hospital testing rates to population testing rates. 

\textbf{Population Testing and Positivity Rates} In some analyses we compare hospital testing and positivity to population testing and positivity. Hospital testing and positivity are defined over a week-long span for a given hospitalization. To make the comparison with the general population clean, we examine test rates and positivity in a given week-long period. We say that a person was tested if she was tested at least once in a given week, and we say she was positive if she was positive at least once in that period.

\subsection{Sample construction}

Throughout, a patient is in the ``test sample" if they are tested at least once. We say a patient is in the ``inpatient sample" if they are hospitalized at least once. We limit our analysis to admissions with non-missing diagnostic information, and we say a patient is in the ``inpatient diagnoses sample" if they meet this restriction. This limitation is important because diagnostic information is necessary for distinguishing COVID-related admissions from non-COVID-related admissions.

\textbf{ICLI and non-ICLI Hospitalizations} We construct three analytic samples from the inpatient data. We start by defining  hospitalizations for influenza- and COVID-like illness (ICLI) using ICD-10 codes. We identify admissions with any of a standard set of ICD-10 codes for ICLI following \citet{afhsc2015}. Then we identify admissions with any of the additional ICD-10 codes that the CDC recommends using for coding COVID hospitalizations \citep{cdcCovidCodes}. Both the influenza-like and COVID-like diagnoses include general symptoms such as cough or fever, as well as more specific diagnoses like acute pneumonia, viral influenza, or COVID-19.  We classify hospitalizations as ICLI-related if they have  any influenza- or COVID-like illness (ICLI) diagnoses, and we classify hospitalizations as non-ICLI if they are not ICLI-related. \hyperref[app:cause]{Appendix \ref*{app:cause}} lists the ICD-10 codes used to define the analytic samples. ICLI hospitalizations may contain diagnostic codes for other (non-ICLI) illnesses. A given hospitalization cannot be both ICLI and non-ICLI, but a given patient with multiple hospitalizations can have ICLI and non-ICLI hospitalizations.%
\footnote{We  view it as an advantage of this definition that a given patient can have a  non-ICLI and an ICLI hospitalization, because we want our hospital-based bounds to count initially mild COVID-19 cases that nonetheless develop into serious cases. For example, a patient hospitalized for labor and delivery with asymptomatic COVID that later develops into a serious case, requiring an ICLI hospitalization, should be counted as an initial negative but a subsequent positive. Our definition counts this patient properly.}

We view the non-ICLI sample as a useful starting point for our analysis for two reasons. First, our hospital IV assumptions are most plausible for hospitalizations that are not obviously COVID-related, and this sample meets that criteria. Second, as we have noted, many states already classify hospitalizations as ICLI-related; thus non-ICLI hospitalizations are identifiable and measurable in near-real time, so this sample can be studied more broadly. 

We acknowledge, however, that the non-ICLI sample may not satisfy the hospital IV assumptions for at least two reasons. First, it may condition on COVID itself, since a patient with a reported COVID diagnosis would be excluded from it. (In practice we observe many patients with positive COVID tests but no COVID diagnosis.)  Second, COVID is a new disease with heterogeneous symptoms, so even if a patient is hospitalized because of COVID, she may not have one of our flagged diagnoses, and we may incorrectly call her hospitalization non-ICLI \citep{YangEtAl2020}.

\textbf{Clear-cause sample} To avoid these problems, we study a third sample, which we call the ``clear cause" sample. These are hospitalizations with a clear cause that is not obviously COVID-related. We define clear-cause hospitalizations as hospitalizations with a diagnosis code for labor and delivery, AMI, stroke, fractures, crushes, open wounds, appendicitis, vehicle accidents, other accidents, or cancer. For all of these conditions except cancer, we flag hospitalizations with a diagnosis at any priority. For cancer, we flag hospitalizations with a cancer diagnosis code as the admitting diagnosis, the primary final diagnosis, or any chemotherapy diagnosis. Note that we do not include among the clear causes respiratory disorders or other diagnoses contributing to our ICLI measure. However admissions can have many diagnosis codes, so it is possible for a clear cause hospitalization to be an ICLI hospitalization.

We view the clear-cause sample as important for two reasons. First, we believe the hospital IV assumptions are most plausible for this sample, so we believe the bounds on prevalence are most likely to be valid. Second, we view the clear-cause sample as offering a test of the validity of the non-ICLI sample. To the extent that the two samples generate similar bounds, we can be more confident that the non-ICLI sample is informative of broader population COVID prevalence, despite the problems with the non-ICLI classification. This would be valuable because classifying hospitalizations as ICLI-related or not requires less information than ascertaining a clear cause of the hospitalization. 

\textbf{Summary statistics} We show summary statistics for all of our samples in \hyperref[tab:ss]{Table \ref*{tab:ss}}, as well as for the state as a whole (from  Census Fact Finder and \cite{inCQF2019}). The average tested and hospitalized patient is substantially older than the population as a whole, and also more likely to be female. Because the tested and hospitalized samples are not age representative of the general population, in what follows we reweight all samples to match the population age distribution.%
\footnote{Specifically, we calculate test rates and positivity rates in week-by-age-group cells, for age groups 0-17, 18-30, 30-50, 50-64, 65-74, and 75 and older. Then we average these age-specific rates across the age groups, weighting each group by its population share.}
The tested and hospitalized samples are fairly similar to the general population in terms of racial composition. Limiting the inpatient sample to admissions with diagnoses reduces our sample size substantially, but it does not appear to change its demographic profile. About one-in-three Hoosiers has ever had a COVID test, whereas about half of hospitalized Hoosiers have had a test. 

\begin{figure}

    \caption{Weekly test rates by sample}
    \label{fig:test_rates}
    \begin{center}
    \includegraphics[width=\textwidth]{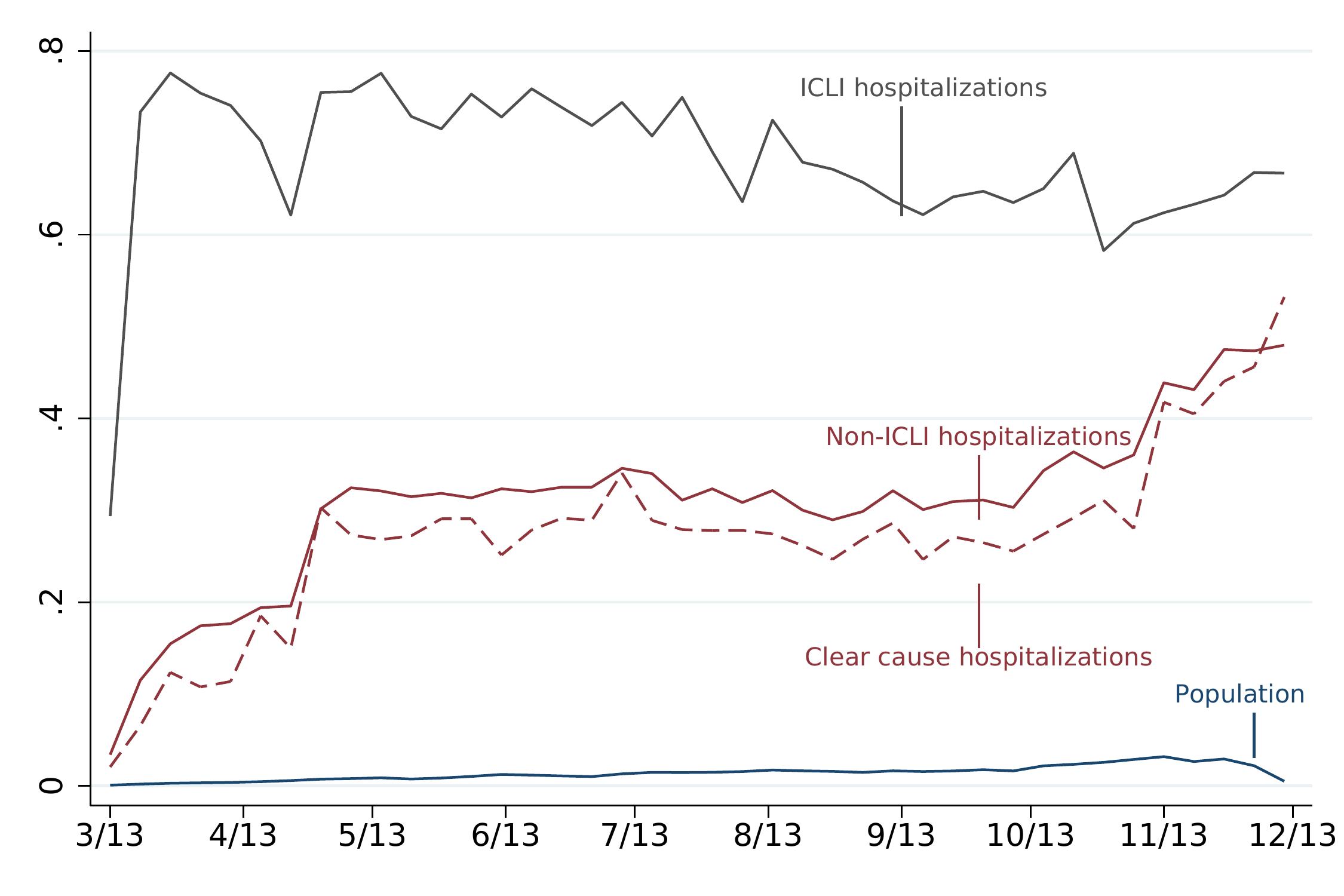}
    \end{center}
    \tabnote{\textwidth}{Notes: Figure plots the age-standardized test rate in each seven-day period of our data, for four samples: the general population, ICLI hospitalizations, non-ICLI hospitalizations, and clear cause hospitalizations. ICLI hospitalizations have at least one diagnosis for influenza-like or COVID-like illness. Clear cause hospitalizations are hospitalizations for cancer, labor and delivery, AMI, stroke, fracture or crush, open wound, appendicitis, or accidents (vehicle or other).  See \hyperref[app:cause]{Appendix \ref*{app:cause}} for definitions. For the general population, the test rate is the fraction of people tested at least once in that week. For the hospitalizations, the test rate is the fraction of hospitalizations admitted in that week with a test between two days prior to admission and four days after. To age-standardize we reweighs the hospitalization samples to match the population age distribution.}
\end{figure}

Although hospitalized patients are about 44 percent (i.e. 49\%/34\%) more likely to have ever been tested than the general public, during the period of their actual hospitalization they are vastly more likely to be tested, as we show in \hyperref[fig:test_rates]{Figure \ref*{fig:test_rates}}. The figure plots age-adjusted weekly testing rates for the whole population and each of our hospitalized samples.%
\footnote{We report the exact values of each of the test rates and the weekly number of admissions in \hyperref[tab:test_rates]{Appendix Table \ref*{tab:test_rates}}. We report age-unweighted test rates in \hyperref[tab:test_rates_unwtd]{Appendix Table \ref*{tab:test_rates_unwtd}}.}. The testing rate in the general population grew from 0.2 percent in April to between 0.4 and 0.6 percent in May and June, and it peaked at about 3 percent in mid-November. So despite increasing more than 10 fold, the weekly test in the Indiana population rate remained below 5 percent for the entire period covered by our study. In contrast, people hospitalized for ICLI were tested at a very high rate, between 60 and 75 percent in most weeks. Testing rates among non-ICLI hospital patients and among the clear-cause non-COVID hospital patients were lower than the ICLI sample but much higher than the population overall, 25-40 percent in May and later months. Testing rates among non-COVID hospital inpatients are 10-25 times higher than testing rates in the general population, but they are typically less than half as high as testing rates in the ICLI population. Despite their very high test rates, hospitalisations are sufficiently rare that hospitalized patients account for a low share of overall tested population, about 4.5 percent in the typical week.

\hyperref[fig:test_rates]{Figure \ref*{fig:test_rates}} shows that although hospitalized patients are tested more often than the general population, they are not always tested. ICLI patients are tested only about two-thirds of the times, and non-ICLI patients only about a third of the time. Several factors contribute to observed tests rates. Highly symptomatic patients may not be tested because a test would not necessarily influence care, and could generate a false negative, and testing capacity was sometimes limited. They also might not receive a SARS-CoV-2 test if they had a positive influenza test, as that provides an alternative explanation for the symptoms.  For symptomatic patients, hospital policy seems to be to encourage testing, but not always require it. The Chief Medical Officer of one large hospital system in the state indicted that asymptomatic patients would typically be tested at admission, but this might vary across hospitals depending on their capacity to isolate patients in private or semi-private rooms \citep{weaver2020}. Another Chief Medical Officer of a large hospital system other reported that testing was at times based on capacity, but patients coming into particular divisions were more likely to be tested, as were patients coming in for operations \citep{crabb2020}. Our personal experience was that hospitals encouraged testing but did not strictly require it.%
\footnote{One of us had a child born in August at a hospital in our sample. The mother was encouraged to obtain a SARS-CoV-2 test prior to admission, but the father (who attended the birth) was not. The mother's test result was not available until after admission (we are happy to report it was negative) and the hospital did not require a rapid test.} Overall we view this anecdotal evidence as indicating that non-ICLI hospitalizations provides a strong encouragement but not a mandate for testing among asymptomatic or mildly symptomatic patients. Greater testing of mildly symptomatic patients would be consistent with our test monotonicity assumption applied to non-ICLI hospitalizations.%
\footnote{Interestingly, test monotonicity might not hold for ICLI patients, as testing might be redundant for the most symptomatic patients. Clinicians indicated that this is more likely in an emergency room context.}

    \section{Bounds on COVID-19 prevalence} 
        \label{sec:bounds}
        \subsection{Bounds by broad samples}

The high test rates among the hospitalized populations shown in \hyperref[fig:test_rates]{Figure \ref*{fig:test_rates}}  imply tight bounds on population prevalence under our monotonicity assumptions. We plot these bounds in \hyperref[fig:bounds]{Figure \ref*{fig:bounds}} and report them in \hyperref[tab:bounds]{Appendix Tables \ref*{tab:bounds} and \ref*{tab:bounds2}}. These bounds are age-adjusted using the same age-group weighting scheme we used for test rates. We report unweighted bounds in \hyperref[tab:bounds_unwtd]{Appendix Tables \ref*{tab:bounds_unwtd} and \ref*{tab:bounds_unwtd2}}. Each panel in \hyperref[fig:bounds]{Figure \ref*{fig:bounds}} plots the test monotonicity bounds for one of our three hospitalizations samples, along with the test monotonicity bounds in the overall population based on the test data. The dashed lines are 95\% confidence intervals around the estimated upper and lower bounds.  (The estimates in the population test data are precise enough that the confidence intervals are indistinguishable from the bounds.)

Several patterns are clear in the figure. First, the ICLI hospitalized population has higher upper and lower bounds on prevalence than the other groups. For the ICLI patients, the prevalence bounds begin at 6-18 percent in the first week of our sample,  increase to 33-39 percent in the last week of March, decline steadily to roughly 12-18 percent in the summer and 8-15 percent in early fall, and increase dramatically in November. Although high, these bounds rule out the possibility that even a majority symptomatic patients are infected with SARS-CoV-2 in any week. In all weeks weeks the ICLI bounds lie outside the other groups' bounds, implying unambiguously higher COVID prevalence among patients hospitalized with influenza- or COVID-like illness. This unsurprising separation shows that the data are sensible and that the bounds are at least informative enough to tell apart these highly distinct populations. 

The second clear pattern in the figure is that the test monotonicity prevalence bounds are much tighter for the non-ICLI and clear-cause hospitalization samples than for the all-test sample. In fact the bounds for both of these hospitalizations samples are always contained within the population-based bounds. At their tightest, the bounds for the all-test sample are as wide as 0.05 to 4.5 percent, for the week of June, 12. In that week the bounds for non-ICLI hospitalizations are [0.7\%, 2.2\%] and for clear-cause hospitalizations they are [0.5\%, 1.8\%]. These tight bounds imply that the hospitalization data could substantially reduce uncertainty about population prevalence under stronger assumptions like hospitalization monotonicity or hospitalization independence.

Third, the bounds for the non-ICLI hospitalization sample and for the clear-cause hospitalization sample are nearly indistinguishable. The only noticeable difference is that the upper bound for non-ICLI hospitalization is perhaps slightly higher. This fact is important because non-ICLI hospitalizations are potentially easier to measure, but they may be negatively selected in the sense that by construction they may exclude COVID-likely cases. The similarity of the non-ICLI bounds with the bounds for the clear-cause sample (which is not selected based on COVID-likelihood) provides some evidence in support of using non-ICLI hospitalizations to measure general prevalence. 

The upper bound for all samples shows a U-shaped pattern, with lower and upper bounds high in the spring, falling in the summer, and rising rapidly in the fall. This pattern does not necessarily indicate that prevalence follows a U-shaped trend, because the lower bound for the population as a whole remains fixed at essentially zero. However the non-ICLI hospitalization bounds are sufficiently tight to confirm that prevalence is lower in mid-summer than late fall. For example, the upper bound the week of September 18 is 1.6 percent; this is lower than the lower bound in any week after October 30. Thus under our monotonicity assumptions, our hospital based bounds are tight enough to show that prevalence unambiguously rose from summer to fall. This rise in prevalence is also evident in mortality data \citep{inCovidDashboard}, but the mortality data show it only with a lag, so the results here show that combining hospital and testing data can be useful for high-frequency monitoring of a pandemic.

Finally, the figures show that although the hospital-based bounds are estimated with less precision than the population-based bounds, the estimates are nonetheless precise enough that in many weeks the confidence intervals remain inside the population bounds. This is especially true for the non-ICLI hospitalizations, which are more common than the ``clear cause" hospitalization. Thus while there is a precision trade-off arising from using a smaller sample with a higher test rate, this trade-off favors the non-ICLI hospitalizations, at least for our data.

\subsection{Bounds by cause of admission}
\label{sec:bounds_cause_admission}

Our overall clear-cause hospitalization sample pools many distinct causes, including among others labor and delivery, vehicle accidents, and other accidents, including falls. In principle these hospitalizations may differ in their SARS-CoV-2 infection risk. One might worry, for example, that pregnant women are especially cautious and careful not to become infected. In contrast, people who get into vehicle accidents during the epidemic might be a less cautious group either because they are not careful drivers, or because they are out of the house at all. 

Since the credibility of key assumptions may vary across different clear causes, we estimated test rates and bounds separately for each of our clear causes of hospitalizations. Because each individual cause has relatively few hospitalizations, we aggregate across all time periods to form these estimates. We focus on nine sets of causes: AMI (i.e. heart attack), appendicitis, cancer, fractures, labor/delivery, non-vehicles accidents, stroke, vehicle accidents, and wounds. These six groups have between 2,000 and 14,000 hospitalizations each. The age profile varies considerably across groups, as we show in \hyperref[tab:ss_group]{Appendix Table \ref*{tab:ss_group}} for age profiles of admitted patients by cause of admission. All ages are represented in the cancer sample. Appendicitis and vehicle accidents both afflict more young people. AMI, stroke, and other accidents---primarily falls---afflict older people; and labor and delivery is limited, of course, to women of childbearing age.%
\footnote{These groups are not necessarily mutually exclusive, and in particular there is overlap between injury and accidents.}
Because not all age groups are represented in every category, we do not age-weight these results.

We report bounds by clear cause of admission in \hyperref[tab:bounds_cause]{Table \ref*{tab:bounds_cause}}. Labor and delivery is tested at the lowest rate, about 20 percent; the other groups are tested 25-40 percent of the time. The bounds are similar across all groups; the (time-pooled) lower bound ranges from .6 percent for vehicle accidents to 3.2 percent for AMI, and the upper bound ranges from 2.1 percent for cancer to 8.7 percent for other accidents. These bounds therefore separate slightly; cancer and vehicle accident stand out as low-prevalence admission types, whereas AMI and other accidents are high prevalence types.  However given the small sample sizes, some amount of separation might be expected, and indeed the 95 percent confidence intervals for these bounds overlap. 
This evidence shows that patients admitted to the hospital for different reasons and with different demographic profiles are all nonetheless tested at a high rate and with similar bounds on prevalence.  Indeed, we might have hypothesized ex ante that cancer patients would be particularly cautious and have the lowest prevalence, but vehicle accident patients (who may be younger and/or more likely to work in essential occupations, given the fact that they were involved in vehicle accidents) would have the highest. Instead we see both groups are on the low end. Thus overall there is no clear evidence that prevalence bounds differ substantially across types of admissions. This is perhaps reassuring for the view that pooling many distinct causes of admissions can nonetheless generate meaningful bounds on prevalence.

    \section{Assessing the hospital representativeness assumptions} 
        \label{sec:validation}
        The results so far show that the test monotonicity bounds on prevalence are much tighter for the non-ICLI hospitalized population than for the population as a whole. These tighter bounds are informative for general population prevalence only under additional assumptions about hospital representativeness, either a monotonicity assumption or an equal prevalence assumption. How valid are these assumptions? Assessing them directly is of course impossible because we lack data on prevalence in the population as a whole or in the hospital sample.

We have already provided one type of indirect evidence in support of our hospital representativeness assumptions. The non-ICLI and clear-cause samples generate similar bounds, and, within the clear-cause sample, there are not large differences in bounds across different causes of admission. This suggests that prevalence does not vary with the exact set of hospitalizations studied, although of course this does not prove hospitalization monotonicity or hospitalization independence are credible assumptions. 

In this section, we provide two additional pieces of evidence on the hospital IV assumptions. First we show that the hospital bounds are consistent with the estimates of population prevalence from the Indiana COVID-19 Random Sample Study \citep{NirWave1,NirWave2}.%
\footnote{Our data do not contain the test results from the Random Sample Study, so we compare our bounds to the published results.}
Second, we compare the hospital sample to the general population in terms of their likelihood of prior testing (prior to the hospital data) and the test rate of their home counties. We take these to be proxies for their concern about COVID, although other interpretations are possible. 

\subsection{Comparison to random sample testing}

A valuable benchmark for the hospital-based prevalence bounds comes from a large-scale study of SARS-CoV-2 prevalence in Indiana. The study invited a representative sample of Indiana residents (aged 12 and older) to obtain a SARS-CoV-2 test. The first wave of the study took place April 25-29, and the second wave took place June 3-7. The preliminary results are reported in \cite{NirWave1} and \cite{NirWave2}. The response rate was roughly 25 percent, and no attempt was made to correct for non-random response. Nonetheless this survey appears to be the best benchmark available. We report the point estimates for prevalence (assuming random nonresponse) and their confidence intervals in the top panel of \hyperref[tab:pop_prev]{Table \ref*{tab:pop_prev}}. The first wave estimates 1.7 percent prevalence and the second 0.5 percent.%
\footnote{The estimates in \hyperref[tab:pop_prev]{Table \ref*{tab:pop_prev}} are slightly different from those reported by \cite{NirWave2}. We report updated calculations, based on correspondence with the authors.}

We compare our prevalence bound during the same time periods in the bottom panel of the table. We limit our sample to tests of people aged 12 and older, for comparison with the population study. Using population testing we obtain very wide bounds that contain the random sample study estimates. This fact provides some support for the test monotonicity assumption. For both the non-ICLI hospitalization and clear-cause hospitalization samples, the bounds are much tighter and the point estimate from the random sample survey lies near the bottom of the identification region in both samples. The confidence interval around upper and lower bounds includes the point estimate from the random sample survey in both April and June for both the non-ICLI and clear-cause samples. Thus for both dates the prevalence point estimates are consistent with the bounds obtained from the non-COVID hospitalizations. As a comparison we also report the bounds from the ICLI-related hospitalizations, which always exclude the random sample estimates.

\subsection{Comparison of prior testing and community testing}
A standard way of measuring representativeness is to compare the distribution of covariates in a study population to their distribution in the target population. In our case, this approach is most convincing if we have well-measured covariates that proxy for SARS-CoV-2 infection risk. Two candidate covariates are the community SARS-CoV-2 testing rate and the prior testing rate. The idea behind these proxies is that people who come from areas with high test rates, or who have been tested in the past, may themselves have a higher current likelihood of being infected with the virus.

To operationalize these measures, we define the community testing rate for person $i$ as the fraction of people in $i$'s county who have ever been tested, as of the end of our sample period. We define the prior test rate of person $i$ as of date $t$ as the probability that $i$ was tested at least once during the week-long period $[t-15, t-9]$. We focus on this window because it is the second week prior to our hospital testing window (which runs from $t-2$ to $t+4$ for a patient admitted at $t$). We allow for a week of time to elapse between the hospitalization and the ``prior" testing because it is possible that some pre-hospital testing would occur in the window $[t-8, t-3]$. When studying prior tests, we limit the sample to each person's first hospitalization after March 1, 2020, to avoid picking up the higher testing that mechanically results from the fact that people hospitalized once are more likely than the general population to have been previously hospitalized. As with our bounds, we weight the data to match the population age distribution. 

\hyperref[tab:county_rates]{Table \ref*{tab:county_rates}} shows the community testing rate. The average county in Indiana has a testing rate of 25\%, with an interquartile range of 22\% to 28\%. The average person lives in a county with a test rate of 26.7\%. The average non-ICLI hospitalized patient comes from a county with a test rate of 26.6\%, and the average clear-cause hospitalization patient comes from a county with a test rate of 26.5\%. Among ICLI hospitalizations it is 26.7\%. Our sample size is large enough that these differences are all statistically significant. Practically, however, the differences are very small. Hospitalized patients appear to come from counties that are roughly representative in terms of their testing rates. These rates are all significantly different from the population average. 

\begin{figure}

    \caption{Prior test rates, population and hospitalization samples }
    \label{fig:prior_rates}
    \begin{center}
    \includegraphics[width=\textwidth]{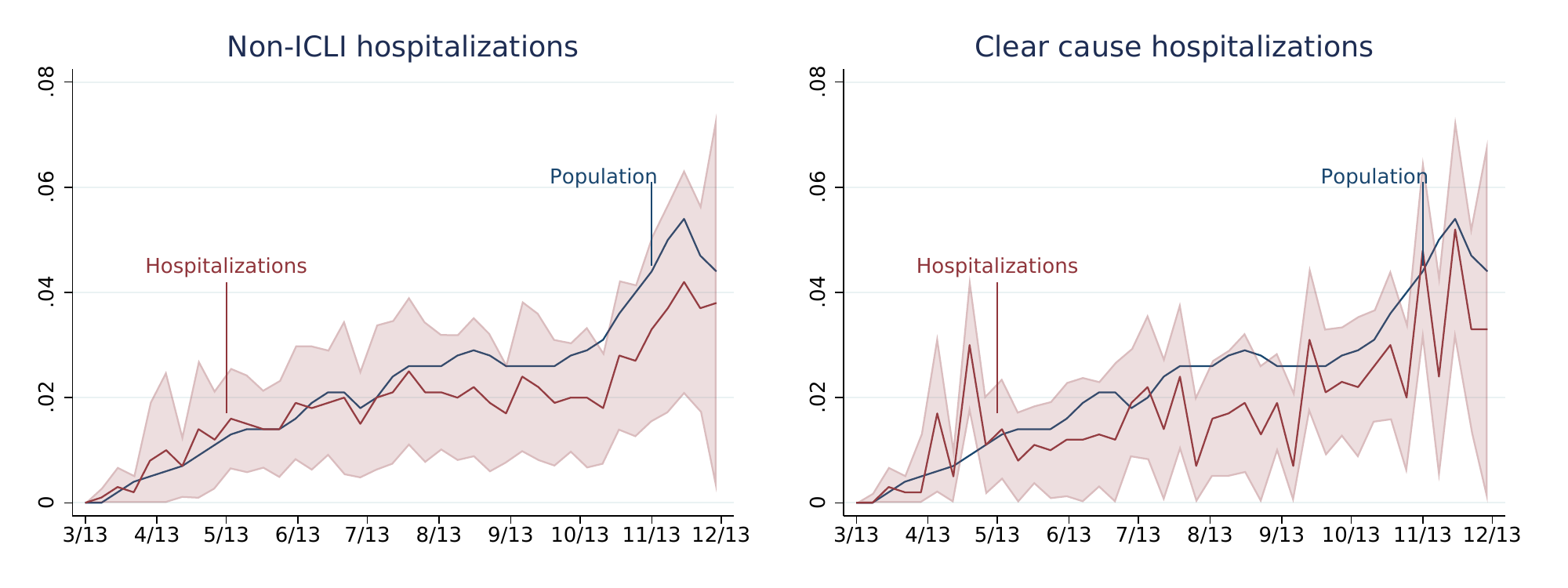}
    \end{center}
    \tabnote{\textwidth}{Notes: The prior test rate is the fraction of the group at date $t$ that was tested between $t-15$ and $t-9$. Figure plots the average prior test rate for the population, for non-ICLI hospitalizations (in the left panel) and for clear cause hospitalizations (right panel). The shaded area is the 95\% confidence interval for each week and hospitalization sample.}
\end{figure}

\hyperref[fig:prior_rates]{Figure \ref*{fig:prior_rates}} shows the prior testing rate as a function of admission date for the non-ICLI hospitalization sample, the clear-cause hospitalization sample, and the general population (for which the prior test rate on day $t$ is defined as the fraction tested between $t-15$ and $t-9$). The rates in the hospitalization samples are initially close to the population rate (when testing is low in general), but the lines diverge. By the last week of the sample, the prior testing rate is 1-2 percentage points lower in the hospitalization samples, than in the population.  Although the differences in weekly testing rates are not statistically significant, the lower prior testing rate in the hospital sample could indicate that the hospital sample is negatively selected on SARS-CoV-2 infection risk.

	\section{Conclusion} 
	    \label{sec:conclusion}
	    We have calculated weekly bounds on the prevalence of SARS-CoV-2 for the Indiana population as a whole and for three hospitalized populations: people hospitalized for influenza- and COVID-like illness, people hospitalized for other reasons not related to influenza or COVID, and people with clear (and clearly not COVID) causes of hospitalization. The bounds we report are valid under weak test monotonicity assumptions. The bounds for the general population are wide but narrow over time. The bounds for the hospitalized population are much tighter, because the hospitalized populations are tested at a much higher rate than the general population.

The hospitalized populations are informative for the general population only under additional, stronger assumptions. In particular, if the hospitalized population is representative of the general population in terms of SARS-CoV-2 prevalence, then both the upper and lower bounds are valid. If the hospitalized population has a higher prevalence  than the general population, then only the upper bound is valid. We assess these assumptions in multiple ways. We find that the confidence intervals around our non-COVID hospitalized population bounds contained the point estimates of prevalence from random sample testing surveys conducted in late April and early June. Since the summer months, hospitalized patients also appear to have lower prior testing rates than the general population, even outside the hospital. The gap in prior testing rates in recent months is not statistically significant, but the point estimates suggest that the hospitalization monotonicity and hospitalization independence assumption could be violated, although the magnitude may not be too severe. Overall, the two hospital representativeness assumptions we consider appear to be plausible, and they have substantial identifying power.  Even under the weak hospitalization monotonicity condition, the hospital data bringing down the upper bound on population prevalence by a third or more.

The basic strategy we pursue in this paper is built on the observation that testing rates differ substantially in the general population and the hospitalized population. The difference in testing rates means that assumptions that permit limited forms of extrapolation from the high testing rate hospitalized population to the general population can substantially reduce uncertainty about the overall prevalence of SARS-CoV-2, which is an important measure of population health. Although we focus on the hospitalized population, our basic strategy might also be applicable to other sub-populations that are tested at higher than usual rates, such as health care workers or university students. 

We believe that the main promise of the non-COVID hospitalization population is that it  can provide near-real time information about population prevalence. Only three numbers are necessary to calculate bounds from the non-COVID hospitalizations: the count of non-COVID admissions, the number of tests among this group, and the number of positive results. Although these numbers are not currently reported, many states already report related numbers, including both the number of COVID tests and the count of ICLI-related hospitalizations. Thus the infrastructure largely exists already to calculate these bounds. The results here help validate this approach for real-time surveillance. 
	    
	\newpage
	
	\bibliographystyle{chicago}
    \bibliography{7.references.bib}

    \clearpage


\begin{figure}

    \caption{Weekly bounds on prevalence under test monotonicity, by sample }
    \label{fig:bounds}
    \begin{center}
    \includegraphics[height=0.80\textheight]{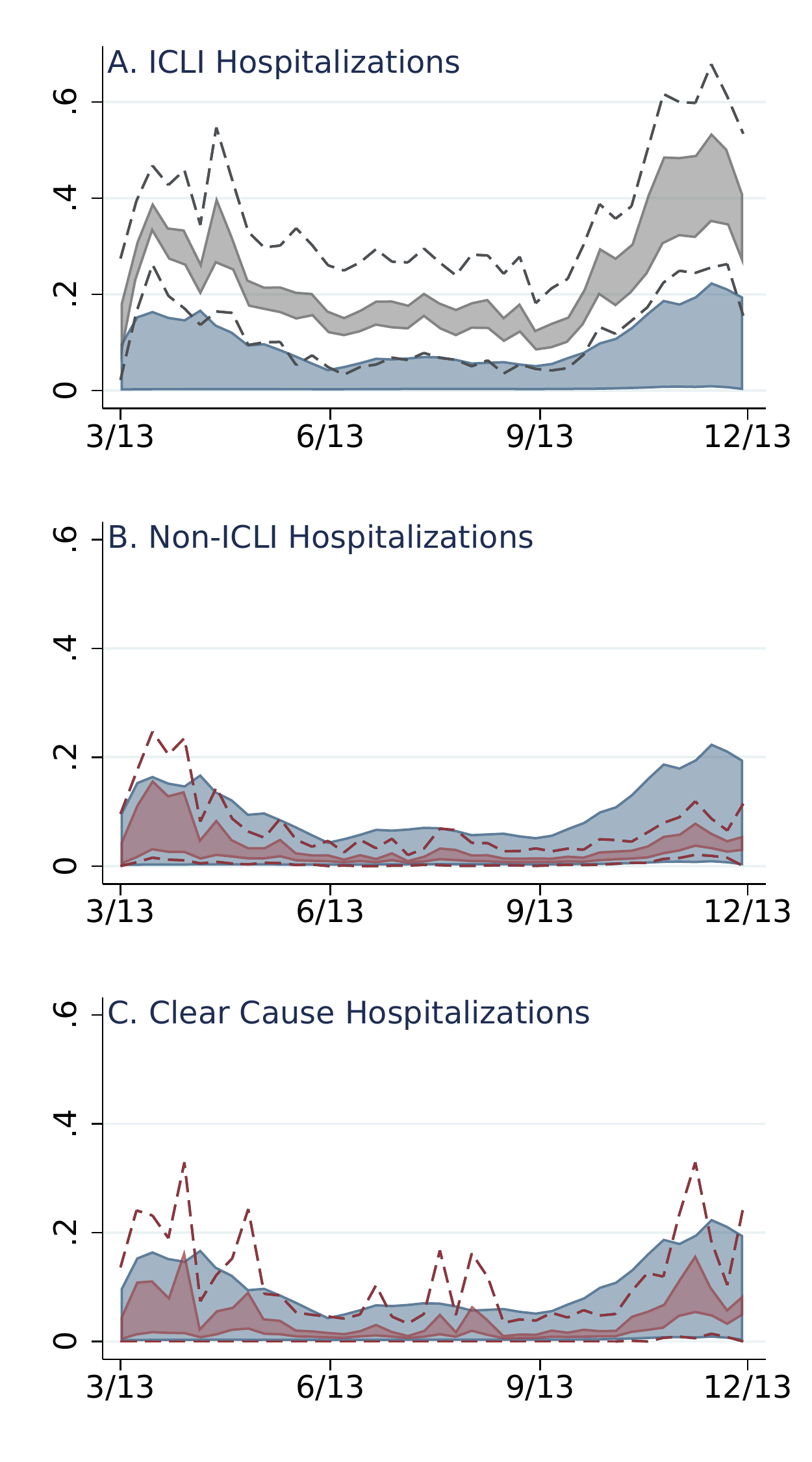}
    \end{center}
    \tabnote{\textwidth}{Notes: Figure plots the age-standardized weekly bounds on prevalence  under test monotonicity, for different samples. The navy area in each plot are the bounds for the general population. The remaining samples are ICLI hospitalizations, non-ICLI hospitalizations, and clear cause hospitalizations. The lower bound is the confirmed positive rate and the upper bound is the positivity rate. ICLI hospitalizations have at least one diagnosis for influenza-like or COVID-like illness. Clear cause hospitalizations are hospitalizations for cancer, labor and delivery, AMI, stroke, fracture or crush, open wound, appendicitis, or accidents (vehicle or other). See \hyperref[app:cause]{Appendix \ref*{app:cause}} for definitions. The dashed line depict pseudo-confidence 95\% confidence intervals. These intervals are tight enough for the general population that they are visually indistinct.  To age-standardize we reweighs the hospitalization samples to match the population age distribution.}
\end{figure}

\clearpage 


\begin{table}

    \caption{Person-level summary statistics}
    
    \label{tab:ss}

    \begin{center}
    \begin{tabular}{l r r rrrrr} \toprule
    \multicolumn{3}{c}{  }  & \multicolumn{5}{c}{Hospitalized}\\
    \cmidrule(lr){4-8}
           & Full  & Ever   &      &    Has    & Not   & Clear \\ 
    Sample & State & Tested & Ever & Diagnosis &  ICLI & cause & ICLI \\ 
    & (1) & (2) & (3) & (4) & (5) & (6) & (7)\\ \midrule
    Age as of 1/1 \\       
Born after 1/1 & & 0.004 &  0.118 &  0.129 &  0.143 &  0.003 &  0.005 \\ 
0-17 & 0.237 & 0.142 &  0.036 &  0.032 &  0.030 &  0.029 &  0.034 \\ 
18-29 & 0.166 & 0.208 &  0.121 &  0.123 &  0.133 &  0.181 &  0.039 \\ 
30-50 & 0.250 & 0.279 &  0.184 &  0.180 &  0.183 &  0.182 &  0.138 \\ 
50-64 & 0.197 & 0.202 &  0.204 &  0.197 &  0.189 &  0.191 &  0.264 \\ 
65-74 & 0.087 & 0.095 &  0.162 &  0.158 &  0.150 &  0.171 &  0.231 \\ 
75+ & 0.063 & 0.070 &  0.175 &  0.182 &  0.172 &  0.243 &  0.289 \\ 
Age unknown & & 0.001 &  0.001 &  0.000 &  0.000 &  0.000 &  0.000 \\ 
\\       
Gender \\       
Male &  0.493 & 0.442 &  0.417 &  0.428 &  0.420 &  0.393 &  0.496 \\ 
Female &  0.507 & 0.558 &  0.583 &  0.572 &  0.580 &  0.607 &  0.504 \\ 
       
Unknown  &  & 0.012 &  0.001 &  0.000 &  0.000 &  0.000 &  0.000 \\ 
\\       
Race/ethnicity \\       
White &  0.848 & 0.877 &  0.862 &  0.834 &  0.835 &  0.853 &  0.828 \\ 
Black &  0.099 & 0.101 &  0.121 &  0.147 &  0.146 &  0.128 &  0.158 \\ 
Race unknown &  & 0.000 &  0.000 &  0.000 &  0.000 &  0.000 &  0.000 \\ 
\\       
Test variables \\       
Ever tested & 0.343 &  1.000 &  0.489 &  0.526 &  0.498 &  0.550 &  0.794 \\ 
Confirmed positive & 0.058 &  0.169 &  0.085 &  0.092 &  0.061 &  0.069 &  0.308 \\ 
\\       
People & 6,637,426 & 2,278,910 &  539,903 &  325,410 &  291,650 &  66,887 &  51,870 \\ 
Counties & 92 & 92 &  92 &  92 &  92 &  92 &  92 \\ 

    \bottomrule
    \end{tabular}
    \end{center} 
    \tabnote{\textwidth}{Notes: Column 1 reports characteristics for the set of people appearing in the test data, and columns 2-6 for people appearing the hospital data, ever (column 2), with at least one diagnosis (column 3), at least one non-ICLI hospitalization for ICLI (column 4), at least one clear cause hospitalization (column 5, see text for details), or  at least one ICLI hospitalization with a diagnosis and not for ICLI (column 6).}
\end{table}

\begin{table}

    \caption{Bounds on prevalence by cause of admission, pooling all time periods}
    \label{tab:bounds_cause}

    \begin{center}
    \begin{tabular}{l rrr} \toprule
    Cause of admission & \# Admissions & Test rate & Bounds \\
    \midrule
    AMI  &  8,624  &  .382  &  [.032, .085] \\
&  &  &  (.028, .094) \\
Appendicitis  &  1,961  &  .383  &  [.017, .045] \\
&  &  &  (.011, .059) \\
Cancer  &  9,586  &  .337  &  [.007, .021] \\
&  &  &  (.005, .026) \\
Fracture  &  13,718  &  .363  &  [.014, .040] \\
&  &  &  (.012, .046) \\
Labor/delivery  &  13,304  &  .197  &  [.008, .040] \\
&  &  &  (.006, .048) \\
Other accident  &  9,782  &  .312  &  [.027, .087] \\
&  &  &  (.024, .097) \\
Stroke  &  8,298  &  .250  &  [.016, .066] \\
&  &  &  (.013, .077) \\
Vehicle accident  &  1,944  &  .271  &  [.006, .024] \\
&  &  &  (.003, .037) \\
Wound  &  3,642  &  .309  &  [.023, .077] \\
&  &  &  (.018, .092) \\

    \bottomrule
    \end{tabular}
    \end{center}
    \tabnote{\textwidth}{Notes: Table reports the number of admissions, in-hospital test rate,  and bounds on COVID prevalence, under test monotonicity, by cause of admission. The sample consists of all admissions with the indicated cause between March 1 and December 18, 2020. See \hyperref[app:cause]{Appendix \ref*{app:cause}} for a precise definition of each cause. We report the point estimate for the bounds in brackets. In parentheses we the lower 2.5th percentile of the lower bound's confidence interval and the upper 97.5th percentile of the upper bound's confidence interval. } 
\end{table}

\begin{table}

    \caption{Do our bounds contain estimates of prevalence from random-sample testing?}
    \label{tab:pop_prev}

    \begin{center}
    \begin{tabular}{l rr} \toprule
    Time period & April 25-29 & June 3 -7 \\ \midrule
    \\
    \underline{Random Sample Study} \\
    Prevalence estimates & 0.0170          & 0.005 \\ 
   95\% confidence interval         & (0.011, 0.025) &  (0.002, 0.013)  \\
    \\
    \underline{Bounds from...} \\ 
    Population testing & [0.0007, 0.138] &  [0.0004, 0.062] \\
&  (0.0006, 0.150) &  (0.0004, 0.069) \\
Non-ICLI hospitalizations & [0.0210, 0.091] &  [0.0093, 0.030] \\
&  (0.0106, 0.136) &  (0.0029, 0.051) \\
Clear cause hospitalizations & [0.0140, 0.068] &  [0.0081, 0.026] \\
&  (0.0000, 0.150) &  (0.0000, 0.066) \\
ICLI-related hospitalizations & [0.3200, 0.477] &  [0.2126, 0.282] \\
&  (0.2054, 0.619) &  (0.0944, 0.442) \\

    \bottomrule
    \end{tabular}
    \end{center}
    \tabnote{\textwidth}{Notes: The first two rows of the table report the estimated population prevalence and 95\% confidence interval from the Indiana COVID-19 Random Sample Study, conducted over the indicated dates, which assumes random nonresponse \citep{NirWave1, NirWave2}. The remaining rows report the (age-adjusted) bounds on prevalence, in brackets, with pseudo 95 confidence intervals, in parentheses, from our different samples. We limit our sample to people aged 12 and older, for consistency with the Random Sample Study.} 
\end{table}

\begin{table}
    
    \caption{Hospitalized patients are not drawn from counties with high test rates}
    \label{tab:county_rates}
    \begin{center}
        \begin{tabular}{l r} \toprule
         & County test rate  \\ \midrule
         Average & 0.252 \\
25th percentile & 0.219 \\
75th percentile & 0.280 \\

         \\
         \underline{Population} \\ Average person &  .267 \\ \\ \underline{Hospitalizations} \\
Non-ICLI  &  .266 \\
&  [20.8] \\
Clear cause  &  .265 \\
&  [16.9] \\
ICLI  &  .267 \\
&  [2.5] \\

         \bottomrule
        \end{tabular}
    \end{center}
    \tabnote{\textwidth}{Notes:  The county test rate is the share of the county population tested at least once in our test data. Table reports county-level statistics, as well as the average county test rates for the general population, the non-ICLI hospitalizations, clear cause hospitalizations, and ICLI hospitalizations, as well as t-statistic (in brackets) for the null hypothesis that the average person and the average hospitalization have the same county test rate.}

\end{table}

	\clearpage
	\appendix

\renewcommand\thetable{\thesection.\arabic{table}}
\renewcommand\thefigure{\thesection.\arabic{figure}}
\renewcommand\theequation{\thesection.\arabic{equation}}
\renewcommand*{\theHtable}{\arabic{section}.\arabic{table}}

\setcounter{table}{0} \renewcommand{\thetable}{A.\arabic{table}}
\setcounter{figure}{0} \renewcommand{\thefigure}{A.\arabic{figure}}

{\flushleft \huge{For online publication}}

	\section{Appendix figures and tables}
	
\begin{figure}[h!]

\caption{Positive cases in our data and on state dashboard}

\label{fig:compare_to_state}

\begin{center}
\includegraphics[width=\textwidth]{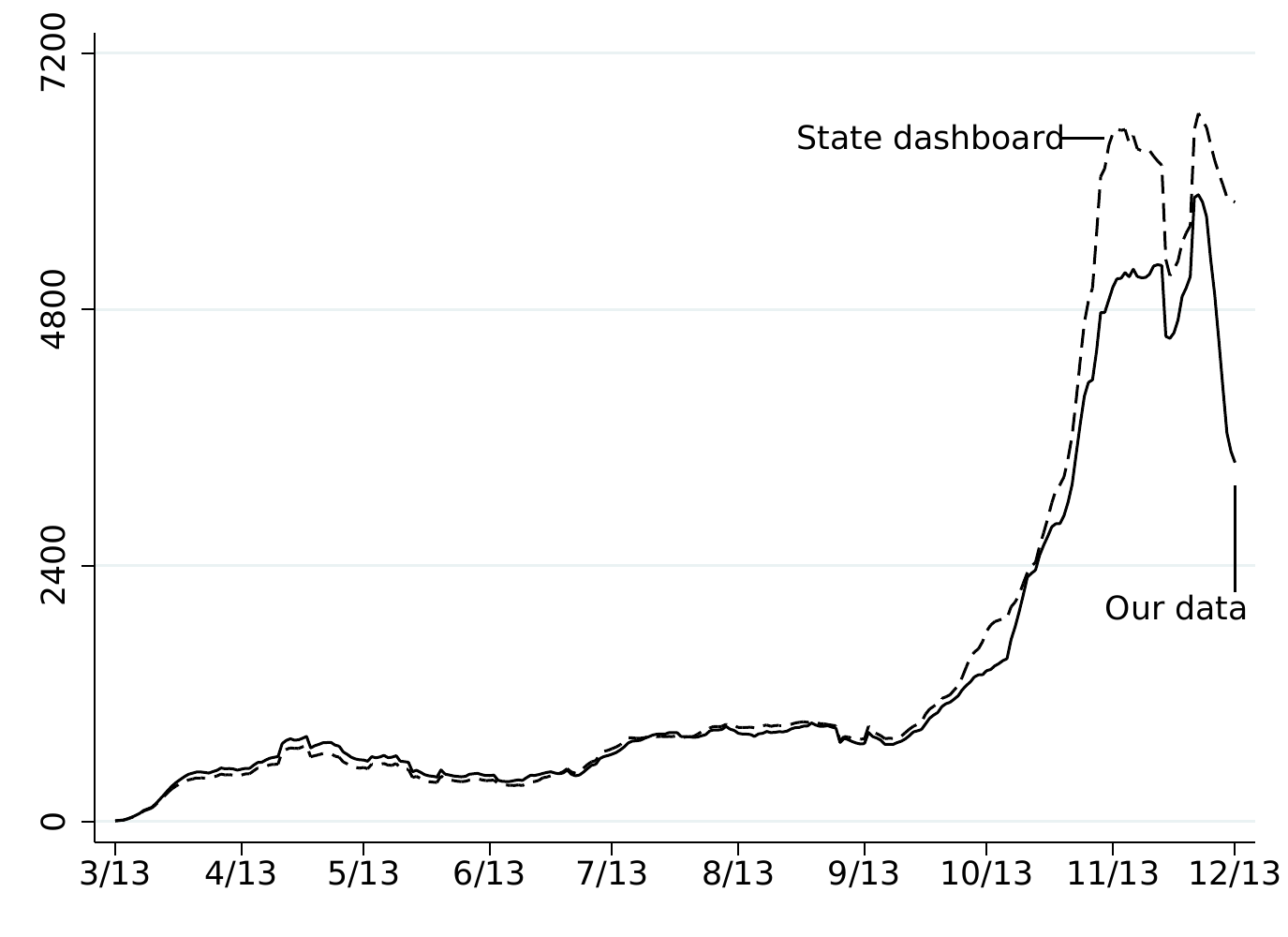}
\end{center}
\tabnote{\textwidth}{Notes: Figure plots 7-day moving average of the number of positive cases reported on Indiana's COVID-19 dashboard \citep{inCovidDashboard}, as well as the number of positive cases observed in our data.}
\end{figure}

\begin{figure}

    \caption{Timing of tests relative to hospitalization}
    \label{fig:hospital_test_times}
    \begin{center}
    \includegraphics[width=\textwidth]{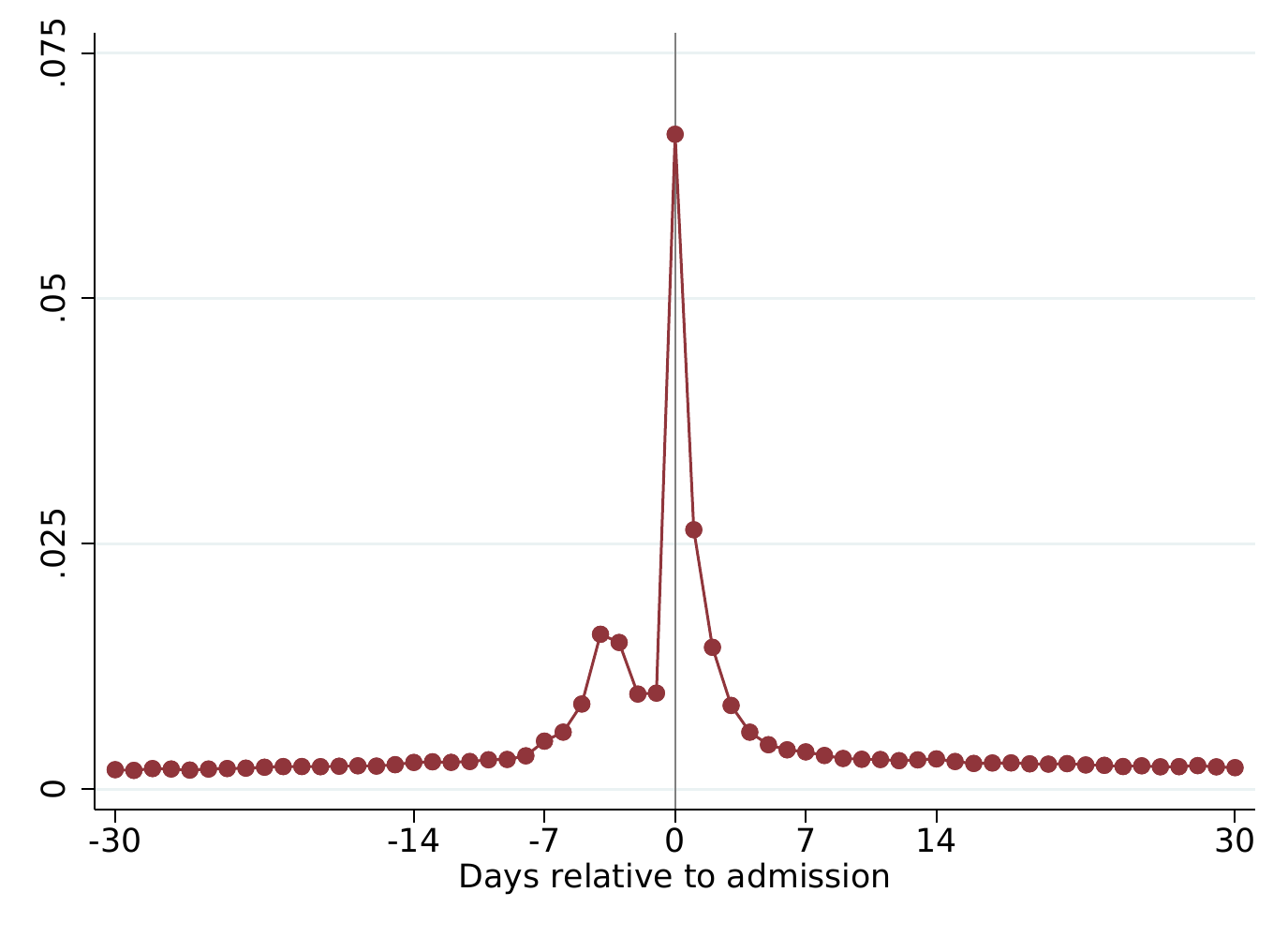}
    \end{center}
    \tabnote{\textwidth}{Notes: Figure plots the fraction of hospitalized patients who had a SARS-CoV-2 test on the indicated day relative to their admission, for non-ICLI hospitalizations, defind as hospitalizations with no diagnosis for influenza-like or COVID-like illness. Patients who are never tested are in the denominator, and a patient can be tested on multiple days.}
\end{figure}

\begin{table}
\footnotesize 

\caption{Weekly test rates, by sample}
\label{tab:test_rates}
\begin{center}
\begin{tabular}{l r rr rr rr } \toprule
Sample  & Population     & \multicolumn{2}{c}{Non-ICLI} 
        & \multicolumn{2}{c}{Clear Cause} & \multicolumn{2}{c}{ICLI} \\ 
\cmidrule(lr){2-2} \cmidrule(lr){3-4} \cmidrule(lr){5-6} \cmidrule(lr){7-8}
 & \% Tested & N & \% Tested  & N & \% Tested  & N & \% Tested \\ 
Week & (1) & (2) & (3) & (4) & (5) & (6) & (7) \\  
\cmidrule(lr){1-1}
\cmidrule(lr){2-2} \cmidrule(lr){3-4} \cmidrule(lr){5-6} \cmidrule(lr){7-8}

  3/13/2020 &  0.001 &  6,094 &  0.034 &  1,363 &  0.021 &  1,140 &  0.294 \\ 
3/20/2020 &  0.002 &  5,173 &  0.115 &  1,278 &  0.065 &  1,226 &  0.734 \\ 
3/27/2020 &  0.003 &  4,697 &  0.155 &  1,189 &  0.124 &  1,342 &  0.776 \\ 
4/03/2020 &  0.003 &  4,677 &  0.174 &  1,273 &  0.108 &  1,204 &  0.754 \\ 
4/10/2020 &  0.004 &  4,841 &  0.177 &  1,248 &  0.114 &  1,078 &  0.741 \\ 
4/17/2020 &  0.005 &  5,057 &  0.194 &  1,348 &  0.185 &  1,191 &  0.702 \\ 
4/24/2020 &  0.006 &  5,303 &  0.196 &  1,431 &  0.150 &  1,108 &  0.621 \\ 
5/01/2020 &  0.007 &  5,832 &  0.302 &  1,442 &  0.303 &  1,195 &  0.755 \\ 
5/08/2020 &  0.008 &  6,187 &  0.325 &  1,555 &  0.273 &  1,106 &  0.756 \\ 
5/15/2020 &  0.009 &  6,774 &  0.321 &  1,577 &  0.268 &  1,120 &  0.776 \\ 
5/22/2020 &  0.007 &  6,793 &  0.315 &  1,581 &  0.272 &  1,059 &  0.729 \\ 
5/29/2020 &  0.008 &  6,949 &  0.318 &  1,563 &  0.291 &  975 &  0.715 \\ 
6/05/2020 &  0.010 &  7,359 &  0.314 &  1,715 &  0.291 &  996 &  0.753 \\ 
6/12/2020 &  0.012 &  7,554 &  0.323 &  1,707 &  0.251 &  974 &  0.728 \\ 
6/19/2020 &  0.012 &  7,437 &  0.320 &  1,644 &  0.279 &  946 &  0.759 \\ 
6/26/2020 &  0.011 &  7,434 &  0.325 &  1,655 &  0.291 &  898 &  0.739 \\ 
7/03/2020 &  0.010 &  7,441 &  0.325 &  1,660 &  0.289 &  937 &  0.719 \\ 
7/10/2020 &  0.013 &  7,591 &  0.346 &  1,702 &  0.340 &  1,050 &  0.744 \\ 
7/17/2020 &  0.015 &  7,650 &  0.340 &  1,658 &  0.289 &  992 &  0.708 \\ 
7/24/2020 &  0.014 &  7,671 &  0.311 &  1,675 &  0.279 &  1,028 &  0.750 \\ 
7/31/2020 &  0.015 &  7,516 &  0.323 &  1,634 &  0.278 &  1,126 &  0.690 \\ 
8/07/2020 &  0.016 &  7,722 &  0.308 &  1,638 &  0.278 &  1,106 &  0.636 \\ 
8/14/2020 &  0.017 &  7,709 &  0.321 &  1,751 &  0.274 &  1,048 &  0.725 \\ 
8/21/2020 &  0.016 &  7,745 &  0.300 &  1,770 &  0.262 &  1,114 &  0.679 \\ 
8/28/2020 &  0.016 &  7,733 &  0.290 &  1,715 &  0.247 &  1,006 &  0.671 \\ 
9/04/2020 &  0.015 &  7,448 &  0.299 &  1,637 &  0.268 &  1,015 &  0.657 \\ 
9/11/2020 &  0.016 &  7,871 &  0.321 &  1,717 &  0.286 &  1,030 &  0.637 \\ 
9/18/2020 &  0.016 &  7,857 &  0.301 &  1,719 &  0.246 &  1,074 &  0.622 \\ 
9/25/2020 &  0.016 &  7,791 &  0.310 &  1,678 &  0.271 &  1,146 &  0.641 \\ 
10/02/2020 &  0.018 &  7,655 &  0.311 &  1,713 &  0.265 &  1,306 &  0.647 \\ 
10/09/2020 &  0.016 &  7,476 &  0.303 &  1,622 &  0.256 &  1,414 &  0.635 \\ 
10/16/2020 &  0.022 &  7,333 &  0.343 &  1,597 &  0.274 &  1,457 &  0.650 \\ 
10/23/2020 &  0.023 &  7,356 &  0.364 &  1,603 &  0.292 &  1,521 &  0.689 \\ 
10/30/2020 &  0.026 &  7,379 &  0.346 &  1,568 &  0.311 &  1,629 &  0.583 \\ 
11/06/2020 &  0.029 &  7,508 &  0.360 &  1,705 &  0.280 &  2,126 &  0.612 \\ 
11/13/2020 &  0.032 &  7,013 &  0.439 &  1,579 &  0.418 &  2,280 &  0.624 \\ 
11/20/2020 &  0.026 &  6,268 &  0.431 &  1,462 &  0.405 &  2,152 &  0.633 \\ 
11/27/2020 &  0.029 &  6,433 &  0.475 &  1,466 &  0.440 &  2,151 &  0.643 \\ 
12/04/2020 &  0.022 &  5,668 &  0.474 &  1,217 &  0.456 &  1,716 &  0.668 \\ 
12/11/2020 &  0.005 &  2,284 &  0.480 &  426 &  0.532 &  689 &  0.667 \\ 

  \bottomrule
\end{tabular}
\end{center}
\tabnote{\textwidth}{Notes: Table reports the weekly test rate for the population, and the number of hospitalizations and test rate, by type of hospitalizations, weighted to match the population age distribution.. (The population size is 6.64 million in all weeks.) ICLI hospitalizations have at least one diagnosis for influenza-like or COVID-like illness. Clear cause hospitalizations are hospitalizations for cancer, labor and delivery, AMI, stroke, fracture or crush, open wound, appendicitis, or accidents (vehicle or other). See \hyperref[app:cause]{Appendix \ref*{app:cause}} for definitions.}
\end{table}

\begin{table}
\footnotesize 

\caption{Weekly test rates, by sample, not age weighted}
\label{tab:test_rates_unwtd}
\begin{center}
\begin{tabular}{l r rr rr rr } \toprule
Sample  & Population     & \multicolumn{2}{c}{Non-ICLI} 
        & \multicolumn{2}{c}{Clear Cause} & \multicolumn{2}{c}{ICLI} \\ 
\cmidrule(lr){2-2} \cmidrule(lr){3-4} \cmidrule(lr){5-6} \cmidrule(lr){7-8}
 & \% Tested & N & \% Tested  & N & \% Tested  & N & \% Tested \\ 
Week & (1) & (2) & (3) & (4) & (5) & (6) & (7) \\  
\cmidrule(lr){1-1}
\cmidrule(lr){2-2} \cmidrule(lr){3-4} \cmidrule(lr){5-6} \cmidrule(lr){7-8}

  3/13/2020 &  0.001 &  6,919 &  0.034 &  1,368 &  0.030 &  1,143 &  0.294 \\ 
3/20/2020 &  0.002 &  6,011 &  0.102 &  1,281 &  0.085 &  1,231 &  0.726 \\ 
3/27/2020 &  0.003 &  5,546 &  0.158 &  1,193 &  0.142 &  1,344 &  0.833 \\ 
4/03/2020 &  0.003 &  5,561 &  0.173 &  1,281 &  0.158 &  1,207 &  0.795 \\ 
4/10/2020 &  0.004 &  5,703 &  0.190 &  1,250 &  0.174 &  1,079 &  0.772 \\ 
4/17/2020 &  0.005 &  5,914 &  0.196 &  1,353 &  0.217 &  1,195 &  0.720 \\ 
4/24/2020 &  0.006 &  6,139 &  0.196 &  1,434 &  0.178 &  1,110 &  0.642 \\ 
5/01/2020 &  0.007 &  6,711 &  0.287 &  1,444 &  0.328 &  1,200 &  0.756 \\ 
5/08/2020 &  0.008 &  7,101 &  0.310 &  1,556 &  0.330 &  1,110 &  0.746 \\ 
5/15/2020 &  0.009 &  7,704 &  0.322 &  1,581 &  0.330 &  1,128 &  0.762 \\ 
5/22/2020 &  0.007 &  7,717 &  0.305 &  1,586 &  0.308 &  1,061 &  0.734 \\ 
5/29/2020 &  0.009 &  7,877 &  0.316 &  1,570 &  0.327 &  982 &  0.737 \\ 
6/05/2020 &  0.010 &  8,270 &  0.321 &  1,721 &  0.317 &  1,001 &  0.744 \\ 
6/12/2020 &  0.012 &  8,432 &  0.318 &  1,711 &  0.310 &  983 &  0.702 \\ 
6/19/2020 &  0.012 &  8,313 &  0.319 &  1,651 &  0.319 &  948 &  0.694 \\ 
6/26/2020 &  0.011 &  8,381 &  0.318 &  1,661 &  0.332 &  903 &  0.708 \\ 
7/03/2020 &  0.010 &  8,367 &  0.311 &  1,668 &  0.321 &  945 &  0.705 \\ 
7/10/2020 &  0.013 &  8,547 &  0.325 &  1,708 &  0.341 &  1,052 &  0.691 \\ 
7/17/2020 &  0.015 &  8,655 &  0.327 &  1,661 &  0.335 &  1,006 &  0.706 \\ 
7/24/2020 &  0.014 &  8,601 &  0.307 &  1,680 &  0.305 &  1,033 &  0.712 \\ 
7/31/2020 &  0.015 &  8,505 &  0.320 &  1,641 &  0.315 &  1,135 &  0.674 \\ 
8/07/2020 &  0.016 &  8,666 &  0.304 &  1,642 &  0.322 &  1,108 &  0.663 \\ 
8/14/2020 &  0.017 &  8,619 &  0.301 &  1,755 &  0.292 &  1,052 &  0.649 \\ 
8/21/2020 &  0.016 &  8,707 &  0.287 &  1,776 &  0.293 &  1,121 &  0.647 \\ 
8/28/2020 &  0.016 &  8,688 &  0.284 &  1,715 &  0.283 &  1,010 &  0.617 \\ 
9/04/2020 &  0.015 &  8,381 &  0.287 &  1,647 &  0.293 &  1,020 &  0.635 \\ 
9/11/2020 &  0.016 &  8,860 &  0.305 &  1,719 &  0.308 &  1,040 &  0.651 \\ 
9/18/2020 &  0.016 &  8,797 &  0.297 &  1,724 &  0.286 &  1,075 &  0.635 \\ 
9/25/2020 &  0.016 &  8,717 &  0.305 &  1,687 &  0.305 &  1,157 &  0.624 \\ 
10/02/2020 &  0.018 &  8,585 &  0.306 &  1,721 &  0.310 &  1,310 &  0.656 \\ 
10/09/2020 &  0.016 &  8,369 &  0.302 &  1,629 &  0.304 &  1,422 &  0.655 \\ 
10/16/2020 &  0.022 &  8,191 &  0.335 &  1,599 &  0.326 &  1,463 &  0.638 \\ 
10/23/2020 &  0.024 &  8,190 &  0.348 &  1,608 &  0.341 &  1,530 &  0.661 \\ 
10/30/2020 &  0.026 &  8,252 &  0.332 &  1,574 &  0.322 &  1,639 &  0.610 \\ 
11/06/2020 &  0.029 &  8,326 &  0.365 &  1,710 &  0.341 &  2,138 &  0.640 \\ 
11/13/2020 &  0.032 &  7,900 &  0.429 &  1,586 &  0.465 &  2,287 &  0.673 \\ 
11/20/2020 &  0.027 &  7,140 &  0.418 &  1,467 &  0.460 &  2,159 &  0.661 \\ 
11/27/2020 &  0.029 &  7,253 &  0.453 &  1,473 &  0.491 &  2,163 &  0.671 \\ 
12/04/2020 &  0.022 &  6,384 &  0.452 &  1,225 &  0.491 &  1,725 &  0.669 \\ 
12/11/2020 &  0.005 &  2,522 &  0.406 &  428 &  0.467 &  689 &  0.597 \\ 

  \bottomrule
\end{tabular}
\end{center}
\tabnote{\textwidth}{Notes: Table reports the weekly test rate for the population, and the number of hospitalizations and test rate, by type of hospitalizations. (The population size is 6.64 million in all weeks.) ICLI hospitalizations have at least one diagnosis for influenza-like or COVID-like illness. Clear cause hospitalizations are hospitalizations for cancer, labor and delivery, AMI, stroke, fracture or crush, open wound, appendicitis, or accidents (vehicle or other). See \hyperref[app:cause]{Appendix \ref*{app:cause}} for definitions.}
\end{table}

\begin{table}
\scriptsize  
\caption{Weekly bounds on prevalence under monotonicity, by sample, March-July}
\label{tab:bounds}
\begin{center}
\begin{tabular}{l rr rr rr rr } \toprule
Sample  & \multicolumn{2}{c}{Population}     & \multicolumn{2}{c}{Non-ICLI} 
        & \multicolumn{2}{c}{Clear Cause} & \multicolumn{2}{c}{ICLI} \\ 
\cmidrule(lr){2-3} \cmidrule(lr){4-5} \cmidrule(lr){6-7} \cmidrule(lr){8-9}
 &    Lower & Upper
 &    Lower & Upper
 &    Lower & Upper
 &    Lower & Upper
  \\ 
Week & (1) & (2) & (3) & (4) & (5) & (6) & (7) & (8)\\ \cmidrule(lr){1-1}
\cmidrule(lr){2-3} \cmidrule(lr){4-5} \cmidrule(lr){6-7} \cmidrule(lr){8-9}
  3/13/2020 &  0.0001 &  0.097 &  0.0019 &  0.043 &  0.0017 &  0.044 &  0.0609 &  0.180 \\ 
& (0.0001,  &  0.118) &  (0.0000,  &  0.096) &  (0.0000,  &  0.136) &  (0.0222,  &  0.275) \\ 
3/20/2020 &  0.0003 &  0.154 &  0.0139 &  0.112 &  0.0107 &  0.110 &  0.2300 &  0.309 \\ 
& (0.0003,  &  0.171) &  (0.0069,  &  0.173) &  (0.0000,  &  0.241) &  (0.1635,  &  0.395) \\ 
3/27/2020 &  0.0006 &  0.166 &  0.0286 &  0.159 &  0.0148 &  0.113 &  0.3281 &  0.392 \\ 
& (0.0005,  &  0.181) &  (0.0154,  &  0.247) &  (0.0000,  &  0.231) &  (0.2621,  &  0.468) \\ 
4/03/2020 &  0.0006 &  0.154 &  0.0238 &  0.131 &  0.0135 &  0.083 &  0.2723 &  0.339 \\ 
& (0.0005,  &  0.169) &  (0.0117,  &  0.205) &  (0.0000,  &  0.190) &  (0.1970,  &  0.427) \\ 
4/10/2020 &  0.0006 &  0.149 &  0.0236 &  0.138 &  0.0132 &  0.169 &  0.2599 &  0.335 \\ 
& (0.0006,  &  0.163) &  (0.0106,  &  0.235) &  (0.0000,  &  0.329) &  (0.1710,  &  0.458) \\ 
4/17/2020 &  0.0008 &  0.169 &  0.0115 &  0.051 &  0.0054 &  0.026 &  0.1980 &  0.268 \\ 
& (0.0008,  &  0.183) &  (0.0049,  &  0.082) &  (0.0000,  &  0.074) &  (0.1366,  &  0.344) \\ 
4/24/2020 &  0.0008 &  0.137 &  0.0182 &  0.086 &  0.0108 &  0.057 &  0.2638 &  0.404 \\ 
& (0.0008,  &  0.148) &  (0.0078,  &  0.144) &  (0.0000,  &  0.122) &  (0.1646,  &  0.547) \\ 
5/01/2020 &  0.0009 &  0.122 &  0.0153 &  0.049 &  0.0192 &  0.064 &  0.2499 &  0.322 \\ 
& (0.0009,  &  0.132) &  (0.0046,  &  0.087) &  (0.0000,  &  0.153) &  (0.1617,  &  0.438) \\ 
5/08/2020 &  0.0008 &  0.096 &  0.0119 &  0.035 &  0.0215 &  0.093 &  0.1744 &  0.231 \\ 
& (0.0007,  &  0.104) &  (0.0030,  &  0.063) &  (0.0000,  &  0.243) &  (0.0949,  &  0.330) \\ 
5/15/2020 &  0.0008 &  0.099 &  0.0120 &  0.035 &  0.0120 &  0.043 &  0.1675 &  0.216 \\ 
& (0.0008,  &  0.106) &  (0.0062,  &  0.052) &  (0.0000,  &  0.088) &  (0.1005,  &  0.298) \\ 
5/22/2020 &  0.0006 &  0.087 &  0.0158 &  0.051 &  0.0110 &  0.040 &  0.1607 &  0.217 \\ 
& (0.0006,  &  0.094) &  (0.0053,  &  0.087) &  (0.0000,  &  0.084) &  (0.1012,  &  0.302) \\ 
5/29/2020 &  0.0006 &  0.073 &  0.0083 &  0.025 &  0.0070 &  0.022 &  0.1474 &  0.206 \\ 
& (0.0005,  &  0.079) &  (0.0020,  &  0.048) &  (0.0000,  &  0.053) &  (0.0532,  &  0.338) \\ 
6/05/2020 &  0.0006 &  0.059 &  0.0073 &  0.022 &  0.0064 &  0.021 &  0.1541 &  0.203 \\ 
& (0.0005,  &  0.064) &  (0.0029,  &  0.036) &  (0.0000,  &  0.049) &  (0.0734,  &  0.303) \\ 
6/12/2020 &  0.0005 &  0.045 &  0.0067 &  0.022 &  0.0053 &  0.018 &  0.1192 &  0.166 \\ 
& (0.0005,  &  0.049) &  (0.0001,  &  0.046) &  (0.0000,  &  0.046) &  (0.0489,  &  0.260) \\ 
6/19/2020 &  0.0006 &  0.051 &  0.0047 &  0.014 &  0.0046 &  0.016 &  0.1129 &  0.153 \\ 
& (0.0005,  &  0.056) &  (0.0011,  &  0.025) &  (0.0000,  &  0.042) &  (0.0333,  &  0.250) \\ 
6/26/2020 &  0.0006 &  0.060 &  0.0070 &  0.022 &  0.0071 &  0.021 &  0.1210 &  0.168 \\ 
& (0.0006,  &  0.064) &  (0.0000,  &  0.049) &  (0.0000,  &  0.050) &  (0.0490,  &  0.266) \\ 
7/03/2020 &  0.0007 &  0.069 &  0.0049 &  0.015 &  0.0088 &  0.033 &  0.1347 &  0.188 \\ 
& (0.0006,  &  0.074) &  (0.0000,  &  0.033) &  (0.0000,  &  0.103) &  (0.0542,  &  0.294) \\ 
7/10/2020 &  0.0009 &  0.067 &  0.0086 &  0.026 &  0.0067 &  0.020 &  0.1292 &  0.188 \\ 
& (0.0008,  &  0.071) &  (0.0006,  &  0.050) &  (0.0000,  &  0.046) &  (0.0687,  &  0.268) \\ 
7/17/2020 &  0.0010 &  0.069 &  0.0041 &  0.011 &  0.0042 &  0.012 &  0.1269 &  0.179 \\ 
& (0.0010,  &  0.073) &  (0.0008,  &  0.020) &  (0.0000,  &  0.033) &  (0.0636,  &  0.267) \\ 
7/24/2020 &  0.0010 &  0.072 &  0.0064 &  0.019 &  0.0065 &  0.021 &  0.1522 &  0.204 \\ 
& (0.0010,  &  0.077) &  (0.0023,  &  0.032) &  (0.0000,  &  0.050) &  (0.0784,  &  0.295) \\ 
7/31/2020 &  0.0011 &  0.072 &  0.0104 &  0.034 &  0.0110 &  0.052 &  0.1270 &  0.183 \\ 
& (0.0010,  &  0.076) &  (0.0017,  &  0.069) &  (0.0000,  &  0.168) &  (0.0681,  &  0.266) \\ 
  
\bottomrule
\end{tabular}
\end{center}
\tabnote{\textwidth}{Notes: Table reports weekly bounds on COVID prevalence in the indicated sample under test monotonicity, weighted to match the population age distribution. The lower bound is the confirmed positive rate and the upper bound is the positivity rate. ICLI hospitalizations have at least one diagnosis for influenza-like or COVID-like illness. Clear cause hospitalizations are hospitalizations for cancer, labor and delivery, AMI, stroke, fracture or crush, open wound, appendicitis, or accidents (vehicle or other). See \hyperref[app:cause]{Appendix \ref*{app:cause}} for definitions. In parentheses brackets we report pseudo 95\% confidence intervals.}
\end{table}

\begin{table}
\scriptsize
\caption{Weekly bounds on prevalence under monotonicity, by sample, August-December}
\label{tab:bounds2}
\begin{center}
\begin{tabular}{l rr rr rr rr } \toprule
Sample  & \multicolumn{2}{c}{Population}     & \multicolumn{2}{c}{Non-ICLI} 
        & \multicolumn{2}{c}{Clear Cause} & \multicolumn{2}{c}{ICLI} \\ 
\cmidrule(lr){2-3} \cmidrule(lr){4-5} \cmidrule(lr){6-7} \cmidrule(lr){8-9}
 &    Lower & Upper
 &    Lower & Upper
 &    Lower & Upper
 &    Lower & Upper
  \\ 
Week & (1) & (2) & (3) & (4) & (5) & (6) & (7) & (8)\\ \cmidrule(lr){1-1}
\cmidrule(lr){2-3} \cmidrule(lr){4-5} \cmidrule(lr){6-7} \cmidrule(lr){8-9}
  8/07/2020 &  0.0010 &  0.067 &  0.0087 &  0.032 &  0.0065 &  0.020 &  0.1127 &  0.170 \\ 
& (0.0010,  &  0.070) &  (0.0004,  &  0.066) &  (0.0000,  &  0.048) &  (0.0639,  &  0.240) \\ 
8/14/2020 &  0.0010 &  0.059 &  0.0070 &  0.022 &  0.0173 &  0.066 &  0.1283 &  0.184 \\ 
& (0.0010,  &  0.062) &  (0.0003,  &  0.043) &  (0.0000,  &  0.162) &  (0.0506,  &  0.283) \\ 
8/21/2020 &  0.0010 &  0.060 &  0.0066 &  0.022 &  0.0098 &  0.041 &  0.1278 &  0.191 \\ 
& (0.0010,  &  0.064) &  (0.0010,  &  0.042) &  (0.0000,  &  0.118) &  (0.0629,  &  0.281) \\ 
8/28/2020 &  0.0010 &  0.062 &  0.0052 &  0.016 &  0.0035 &  0.012 &  0.1006 &  0.154 \\ 
& (0.0010,  &  0.065) &  (0.0016,  &  0.027) &  (0.0000,  &  0.034) &  (0.0357,  &  0.243) \\ 
9/04/2020 &  0.0009 &  0.057 &  0.0049 &  0.016 &  0.0042 &  0.015 &  0.1198 &  0.182 \\ 
& (0.0008,  &  0.060) &  (0.0012,  &  0.028) &  (0.0000,  &  0.041) &  (0.0544,  &  0.278) \\ 
9/11/2020 &  0.0009 &  0.053 &  0.0053 &  0.016 &  0.0043 &  0.014 &  0.0830 &  0.127 \\ 
& (0.0008,  &  0.057) &  (0.0004,  &  0.033) &  (0.0000,  &  0.038) &  (0.0451,  &  0.182) \\ 
9/18/2020 &  0.0009 &  0.058 &  0.0051 &  0.016 &  0.0064 &  0.022 &  0.0879 &  0.141 \\ 
& (0.0009,  &  0.061) &  (0.0015,  &  0.027) &  (0.0000,  &  0.053) &  (0.0420,  &  0.213) \\ 
9/25/2020 &  0.0012 &  0.070 &  0.0064 &  0.019 &  0.0061 &  0.018 &  0.0993 &  0.154 \\ 
& (0.0011,  &  0.074) &  (0.0023,  &  0.032) &  (0.0000,  &  0.044) &  (0.0467,  &  0.232) \\ 
10/02/2020 &  0.0014 &  0.081 &  0.0058 &  0.018 &  0.0065 &  0.024 &  0.1364 &  0.210 \\ 
& (0.0014,  &  0.085) &  (0.0018,  &  0.030) &  (0.0000,  &  0.058) &  (0.0745,  &  0.303) \\ 
10/09/2020 &  0.0017 &  0.101 &  0.0086 &  0.027 &  0.0072 &  0.021 &  0.1975 &  0.297 \\ 
& (0.0016,  &  0.105) &  (0.0026,  &  0.049) &  (0.0000,  &  0.048) &  (0.1323,  &  0.388) \\ 
10/16/2020 &  0.0024 &  0.110 &  0.0103 &  0.029 &  0.0073 &  0.022 &  0.1747 &  0.277 \\ 
& (0.0023,  &  0.114) &  (0.0040,  &  0.048) &  (0.0000,  &  0.050) &  (0.1181,  &  0.357) \\ 
10/23/2020 &  0.0031 &  0.132 &  0.0115 &  0.030 &  0.0154 &  0.047 &  0.2028 &  0.305 \\ 
& (0.0030,  &  0.136) &  (0.0061,  &  0.045) &  (0.0013,  &  0.092) &  (0.1454,  &  0.383) \\ 
10/30/2020 &  0.0042 &  0.162 &  0.0136 &  0.038 &  0.0190 &  0.057 &  0.2422 &  0.406 \\ 
& (0.0041,  &  0.166) &  (0.0059,  &  0.062) &  (0.0000,  &  0.126) &  (0.1733,  &  0.500) \\ 
11/06/2020 &  0.0055 &  0.189 &  0.0213 &  0.056 &  0.0230 &  0.069 &  0.3045 &  0.487 \\ 
& (0.0053,  &  0.193) &  (0.0131,  &  0.079) &  (0.0066,  &  0.119) &  (0.2243,  &  0.616) \\ 
11/13/2020 &  0.0059 &  0.181 &  0.0264 &  0.060 &  0.0450 &  0.115 &  0.3206 &  0.486 \\ 
& (0.0057,  &  0.185) &  (0.0149,  &  0.090) &  (0.0087,  &  0.234) &  (0.2492,  &  0.600) \\ 
11/20/2020 &  0.0052 &  0.196 &  0.0352 &  0.080 &  0.0519 &  0.160 &  0.3169 &  0.490 \\ 
& (0.0051,  &  0.201) &  (0.0208,  &  0.119) &  (0.0059,  &  0.329) &  (0.2448,  &  0.598) \\ 
11/27/2020 &  0.0066 &  0.225 &  0.0306 &  0.062 &  0.0457 &  0.100 &  0.3500 &  0.536 \\ 
& (0.0065,  &  0.230) &  (0.0189,  &  0.087) &  (0.0143,  &  0.184) &  (0.2560,  &  0.678) \\ 
12/04/2020 &  0.0047 &  0.213 &  0.0243 &  0.048 &  0.0299 &  0.060 &  0.3432 &  0.502 \\ 
& (0.0046,  &  0.218) &  (0.0152,  &  0.065) &  (0.0080,  &  0.104) &  (0.2630,  &  0.611) \\ 
12/11/2020 &  0.0010 &  0.195 &  0.0277 &  0.056 &  0.0480 &  0.086 &  0.2589 &  0.408 \\ 
& (0.0009,  &  0.205) &  (0.0000,  &  0.115) &  (0.0000,  &  0.246) &  (0.1559,  &  0.534) \\ 

\bottomrule
\end{tabular}
\end{center}
\tabnote{\textwidth}{Notes: Table reports weekly bounds on COVID prevalence in the indicated sample under test monotonicity, weighted to match the population age distribution. The lower bound is the confirmed positive rate and the upper bound is the positivity rate. ICLI hospitalizations have at least one diagnosis for influenza-like or COVID-like illness. Clear cause hospitalizations are hospitalizations for cancer, labor and delivery, AMI, stroke, fracture or crush, open wound, appendicitis, or accidents (vehicle or other). See \hyperref[app:cause]{Appendix \ref*{app:cause}} for definitions. In parentheses we report pseudo 95\% confidence intervals.}
\end{table}

\begin{table}
\scriptsize 

\caption{Weekly bounds on prevalence under monotonicity, by sample, not age-weighted, March-July}
\label{tab:bounds_unwtd}
\begin{center}
\begin{tabular}{l rr rr rr rr } \toprule
Sample  & \multicolumn{2}{c}{Population}     & \multicolumn{2}{c}{Non-ICLI} 
        & \multicolumn{2}{c}{Clear Cause} & \multicolumn{2}{c}{ICLI} \\ 
\cmidrule(lr){2-3} \cmidrule(lr){4-5} \cmidrule(lr){6-7} \cmidrule(lr){8-9}
 &    Lower & Upper
 &    Lower & Upper
 &    Lower & Upper
 &    Lower & Upper
  \\ 
Week & (1) & (2) & (3) & (4) & (5) & (6) & (7) & (8)\\ \cmidrule(lr){1-1}
\cmidrule(lr){2-3} \cmidrule(lr){4-5} \cmidrule(lr){6-7} \cmidrule(lr){8-9}
  3/13/2020 &  0.0001 &  0.113 &  0.0030 &  0.090 &  0.0022 &  0.073 &  0.0717 &  0.244 \\ 
&  (0.0001,  &  0.122) &  (0.0017,  &  0.126) &  (0.0000,  &  0.153) &  (0.0568,  &  0.290) \\ 
3/20/2020 &  0.0003 &  0.181 &  0.0163 &  0.159 &  0.0148 &  0.174 &  0.2868 &  0.395 \\ 
&  (0.0003,  &  0.187) &  (0.0131,  &  0.188) &  (0.0082,  &  0.246) &  (0.2615,  &  0.427) \\ 
3/27/2020 &  0.0006 &  0.196 &  0.0330 &  0.209 &  0.0210 &  0.147 &  0.4308 &  0.517 \\ 
&  (0.0005,  &  0.202) &  (0.0283,  &  0.236) &  (0.0128,  &  0.200) &  (0.4043,  &  0.546) \\ 
4/03/2020 &  0.0006 &  0.177 &  0.0279 &  0.161 &  0.0258 &  0.163 &  0.3529 &  0.444 \\ 
&  (0.0006,  &  0.183) &  (0.0235,  &  0.184) &  (0.0171,  &  0.214) &  (0.3260,  &  0.476) \\ 
4/10/2020 &  0.0006 &  0.168 &  0.0254 &  0.134 &  0.0184 &  0.106 &  0.3105 &  0.402 \\ 
&  (0.0006,  &  0.173) &  (0.0213,  &  0.154) &  (0.0109,  &  0.146) &  (0.2829,  &  0.435) \\ 
4/17/2020 &  0.0008 &  0.187 &  0.0166 &  0.084 &  0.0126 &  0.058 &  0.2636 &  0.366 \\ 
&  (0.0008,  &  0.191) &  (0.0133,  &  0.100) &  (0.0066,  &  0.085) &  (0.2386,  &  0.398) \\ 
4/24/2020 &  0.0008 &  0.146 &  0.0217 &  0.110 &  0.0153 &  0.086 &  0.2694 &  0.419 \\ 
&  (0.0008,  &  0.149) &  (0.0180,  &  0.128) &  (0.0090,  &  0.121) &  (0.2433,  &  0.456) \\ 
5/01/2020 &  0.0009 &  0.125 &  0.0165 &  0.058 &  0.0187 &  0.057 &  0.2467 &  0.326 \\ 
&  (0.0009,  &  0.128) &  (0.0135,  &  0.068) &  (0.0117,  &  0.078) &  (0.2223,  &  0.357) \\ 
5/08/2020 &  0.0008 &  0.097 &  0.0120 &  0.039 &  0.0154 &  0.047 &  0.1901 &  0.255 \\ 
&  (0.0007,  &  0.099) &  (0.0094,  &  0.047) &  (0.0093,  &  0.065) &  (0.1670,  &  0.285) \\ 
5/15/2020 &  0.0008 &  0.094 &  0.0145 &  0.045 &  0.0190 &  0.057 &  0.1826 &  0.240 \\ 
&  (0.0008,  &  0.096) &  (0.0119,  &  0.053) &  (0.0122,  &  0.077) &  (0.1601,  &  0.268) \\ 
5/22/2020 &  0.0006 &  0.084 &  0.0143 &  0.047 &  0.0164 &  0.053 &  0.1913 &  0.261 \\ 
&  (0.0006,  &  0.086) &  (0.0116,  &  0.055) &  (0.0101,  &  0.073) &  (0.1677,  &  0.291) \\ 
5/29/2020 &  0.0006 &  0.069 &  0.0090 &  0.028 &  0.0089 &  0.027 &  0.1415 &  0.192 \\ 
&  (0.0006,  &  0.071) &  (0.0069,  &  0.035) &  (0.0043,  &  0.041) &  (0.1197,  &  0.221) \\ 
6/05/2020 &  0.0006 &  0.056 &  0.0085 &  0.026 &  0.0087 &  0.027 &  0.1608 &  0.216 \\ 
&  (0.0006,  &  0.058) &  (0.0065,  &  0.032) &  (0.0043,  &  0.041) &  (0.1381,  &  0.246) \\ 
6/12/2020 &  0.0005 &  0.041 &  0.0057 &  0.018 &  0.0070 &  0.023 &  0.1312 &  0.187 \\ 
&  (0.0005,  &  0.042) &  (0.0041,  &  0.023) &  (0.0031,  &  0.035) &  (0.1101,  &  0.216) \\ 
6/19/2020 &  0.0006 &  0.049 &  0.0057 &  0.018 &  0.0048 &  0.015 &  0.1118 &  0.161 \\ 
&  (0.0006,  &  0.050) &  (0.0040,  &  0.023) &  (0.0015,  &  0.026) &  (0.0918,  &  0.189) \\ 
6/26/2020 &  0.0006 &  0.058 &  0.0062 &  0.020 &  0.0102 &  0.031 &  0.1152 &  0.163 \\ 
&  (0.0006,  &  0.060) &  (0.0045,  &  0.025) &  (0.0054,  &  0.045) &  (0.0944,  &  0.191) \\ 
7/03/2020 &  0.0007 &  0.067 &  0.0047 &  0.015 &  0.0084 &  0.026 &  0.1291 &  0.183 \\ 
&  (0.0007,  &  0.069) &  (0.0032,  &  0.020) &  (0.0040,  &  0.040) &  (0.1077,  &  0.213) \\ 
7/10/2020 &  0.0009 &  0.067 &  0.0066 &  0.020 &  0.0082 &  0.024 &  0.1416 &  0.205 \\ 
&  (0.0009,  &  0.069) &  (0.0048,  &  0.025) &  (0.0039,  &  0.037) &  (0.1206,  &  0.234) \\ 
7/17/2020 &  0.0010 &  0.069 &  0.0054 &  0.017 &  0.0066 &  0.020 &  0.1471 &  0.208 \\ 
&  (0.0010,  &  0.071) &  (0.0039,  &  0.021) &  (0.0027,  &  0.031) &  (0.1252,  &  0.238) \\ 
7/24/2020 &  0.0010 &  0.072 &  0.0083 &  0.027 &  0.0119 &  0.039 &  0.1626 &  0.228 \\ 
&  (0.0010,  &  0.073) &  (0.0063,  &  0.033) &  (0.0067,  &  0.056) &  (0.1401,  &  0.259) \\ 
7/31/2020 &  0.0011 &  0.071 &  0.0088 &  0.028 &  0.0110 &  0.035 &  0.1692 &  0.251 \\ 
&  (0.0010,  &  0.073) &  (0.0068,  &  0.034) &  (0.0059,  &  0.051) &  (0.1474,  &  0.282) \\ 
  
\bottomrule
\end{tabular}
\end{center}
\tabnote{\textwidth}{Notes: Table reports weekly bounds on COVID prevalence in the indicated sample under test monotonicity. The lower bound is the confirmed positive rate and the upper bound is the positivity rate. ICLI hospitalizations have at least one diagnosis for influenza-like or COVID-like illness. Clear cause hospitalizations are hospitalizations for cancer, labor and delivery, AMI, stroke, fracture or crush, open wound, appendicitis, or accidents (vehicle or other). See \hyperref[app:cause]{Appendix \ref*{app:cause}} for definitions.In parentheses we report pseudo 95\% confidence intervals.}
\end{table}

\begin{table}
\footnotesize 

\caption{Weekly bounds on prevalence under monotonicity, by sample, not age-weighted, August-December}
\label{tab:bounds_unwtd2}
\begin{center}
\begin{tabular}{l rr rr rr rr } \toprule
Sample  & \multicolumn{2}{c}{Population}     & \multicolumn{2}{c}{Non-ICLI} 
        & \multicolumn{2}{c}{Clear Cause} & \multicolumn{2}{c}{ICLI} \\ 
\cmidrule(lr){2-3} \cmidrule(lr){4-5} \cmidrule(lr){6-7} \cmidrule(lr){8-9}
 &    Lower & Upper
 &    Lower & Upper
 &    Lower & Upper
 &    Lower & Upper
  \\ 
Week & (1) & (2) & (3) & (4) & (5) & (6) & (7) & (8)\\ \cmidrule(lr){1-1}
\cmidrule(lr){2-3} \cmidrule(lr){4-5} \cmidrule(lr){6-7} \cmidrule(lr){8-9}
  8/07/2020 &  0.0010 &  0.067 &  0.0070 &  0.023 &  0.0097 &  0.030 &  0.1606 &  0.242 \\ 
&  (0.0010,  &  0.069) &  (0.0053,  &  0.029) &  (0.0050,  &  0.045) &  (0.1390,  &  0.273) \\ 
8/14/2020 &  0.0010 &  0.061 &  0.0068 &  0.023 &  0.0137 &  0.047 &  0.1454 &  0.224 \\ 
&  (0.0010,  &  0.062) &  (0.0051,  &  0.029) &  (0.0082,  &  0.065) &  (0.1241,  &  0.255) \\ 
8/21/2020 &  0.0010 &  0.064 &  0.0063 &  0.022 &  0.0090 &  0.031 &  0.1490 &  0.230 \\ 
&  (0.0010,  &  0.065) &  (0.0047,  &  0.028) &  (0.0046,  &  0.046) &  (0.1281,  &  0.261) \\ 
8/28/2020 &  0.0010 &  0.066 &  0.0061 &  0.022 &  0.0058 &  0.021 &  0.1238 &  0.201 \\ 
&  (0.0010,  &  0.068) &  (0.0045,  &  0.027) &  (0.0022,  &  0.033) &  (0.1035,  &  0.232) \\ 
9/04/2020 &  0.0009 &  0.060 &  0.0068 &  0.024 &  0.0073 &  0.025 &  0.1304 &  0.205 \\ 
&  (0.0009,  &  0.061) &  (0.0050,  &  0.030) &  (0.0032,  &  0.039) &  (0.1097,  &  0.236) \\ 
9/11/2020 &  0.0009 &  0.055 &  0.0056 &  0.019 &  0.0093 &  0.030 &  0.1337 &  0.205 \\ 
&  (0.0009,  &  0.056) &  (0.0041,  &  0.024) &  (0.0048,  &  0.045) &  (0.1130,  &  0.236) \\ 
9/18/2020 &  0.0009 &  0.059 &  0.0064 &  0.021 &  0.0093 &  0.032 &  0.1321 &  0.208 \\ 
&  (0.0009,  &  0.061) &  (0.0047,  &  0.027) &  (0.0048,  &  0.048) &  (0.1119,  &  0.238) \\ 
9/25/2020 &  0.0012 &  0.072 &  0.0073 &  0.024 &  0.0095 &  0.031 &  0.1340 &  0.215 \\ 
&  (0.0011,  &  0.073) &  (0.0055,  &  0.030) &  (0.0049,  &  0.046) &  (0.1143,  &  0.245) \\ 
10/02/2020 &  0.0015 &  0.082 &  0.0080 &  0.026 &  0.0099 &  0.032 &  0.1840 &  0.280 \\ 
&  (0.0014,  &  0.084) &  (0.0061,  &  0.032) &  (0.0052,  &  0.047) &  (0.1630,  &  0.310) \\ 
10/09/2020 &  0.0017 &  0.102 &  0.0103 &  0.034 &  0.0153 &  0.051 &  0.2342 &  0.357 \\ 
&  (0.0016,  &  0.104) &  (0.0081,  &  0.041) &  (0.0094,  &  0.070) &  (0.2122,  &  0.388) \\ 
10/16/2020 &  0.0024 &  0.111 &  0.0126 &  0.038 &  0.0150 &  0.046 &  0.2502 &  0.392 \\ 
&  (0.0024,  &  0.113) &  (0.0102,  &  0.045) &  (0.0090,  &  0.064) &  (0.2280,  &  0.423) \\ 
10/23/2020 &  0.0031 &  0.134 &  0.0162 &  0.047 &  0.0236 &  0.069 &  0.2876 &  0.435 \\ 
&  (0.0031,  &  0.136) &  (0.0135,  &  0.054) &  (0.0162,  &  0.090) &  (0.2649,  &  0.466) \\ 
10/30/2020 &  0.0042 &  0.163 &  0.0145 &  0.044 &  0.0216 &  0.067 &  0.3130 &  0.513 \\ 
&  (0.0041,  &  0.165) &  (0.0120,  &  0.051) &  (0.0144,  &  0.089) &  (0.2905,  &  0.544) \\ 
11/06/2020 &  0.0055 &  0.191 &  0.0259 &  0.071 &  0.0357 &  0.105 &  0.3728 &  0.582 \\ 
&  (0.0054,  &  0.193) &  (0.0225,  &  0.080) &  (0.0269,  &  0.129) &  (0.3523,  &  0.608) \\ 
11/13/2020 &  0.0059 &  0.184 &  0.0281 &  0.065 &  0.0530 &  0.114 &  0.4215 &  0.626 \\ 
&  (0.0058,  &  0.186) &  (0.0245,  &  0.074) &  (0.0419,  &  0.137) &  (0.4013,  &  0.651) \\ 
11/20/2020 &  0.0053 &  0.198 &  0.0391 &  0.093 &  0.0532 &  0.116 &  0.3983 &  0.602 \\ 
&  (0.0052,  &  0.200) &  (0.0346,  &  0.104) &  (0.0417,  &  0.140) &  (0.3777,  &  0.628) \\ 
11/27/2020 &  0.0066 &  0.227 &  0.0381 &  0.084 &  0.0699 &  0.142 &  0.4170 &  0.621 \\ 
&  (0.0066,  &  0.228) &  (0.0337,  &  0.094) &  (0.0569,  &  0.168) &  (0.3962,  &  0.646) \\ 
12/04/2020 &  0.0047 &  0.214 &  0.0310 &  0.069 &  0.0531 &  0.108 &  0.4261 &  0.637 \\ 
&  (0.0046,  &  0.216) &  (0.0268,  &  0.078) &  (0.0405,  &  0.133) &  (0.4028,  &  0.665) \\ 
12/11/2020 &  0.0010 &  0.196 &  0.0206 &  0.051 &  0.0397 &  0.085 &  0.3570 &  0.599 \\ 
&  (0.0010,  &  0.201) &  (0.0151,  &  0.064) &  (0.0212,  &  0.124) &  (0.3213,  &  0.646) \\ 

\bottomrule
\end{tabular}
\end{center}
\tabnote{\textwidth}{Notes: Table reports weekly bounds on COVID prevalence in the indicated sample under test monotonicity. The lower bound is the confirmed positive rate and the upper bound is the positivity rate. ICLI hospitalizations have at least one diagnosis for influenza-like or COVID-like illness. Clear cause hospitalizations are hospitalizations for cancer, labor and delivery, AMI, stroke, fracture or crush, open wound, appendicitis, or accidents (vehicle or other). See \hyperref[app:cause]{Appendix \ref*{app:cause}} for definitions.In parentheses we report pseudo 95\% confidence intervals.}
\end{table}

\begin{table}
\caption{Demographics and test rates among hospitalized patients, by group}
\label{tab:ss_group}

\begin{center}
\begin{tabular}{l r rrrr rrr} \toprule
    & Number of & \multicolumn{7}{c}{Age} \\ \cmidrule(lr){3-9}
Group & Admissions & Newborn&  0-17 & 18-29 & 30-49 & 50-64 & 65-74  & $>$74 \\ \midrule
All & 781,587 &  0.080 &  0.028 &  0.105 &  0.181 &  0.228 &  0.182 &  0.196 \\ 
Has diagnosis & 355,425 &  0.100 &  0.026 &  0.112 &  0.178 &  0.214 &  0.173 &  0.198 \\ 
ICLI & 49,904 &  0.005 &  0.023 &  0.037 &  0.139 &  0.269 &  0.239 &  0.287 \\ 
Non-ICLI & 305,521 &  0.115 &  0.027 &  0.124 &  0.184 &  0.205 &  0.162 &  0.183 \\ 
Clear cause & 61,684 &  0.003 &  0.033 &  0.165 &  0.175 &  0.195 &  0.179 &  0.249 \\ 
Cancer & 9,586 &  0.001 &  0.053 &  0.026 &  0.122 &  0.322 &  0.284 &  0.192 \\ 
Labor/delivery & 13,304 &  0.009 &  0.023 &  0.611 &  0.357 &  0.000 &  0.000 &  0.000 \\ 
AMI & 8,624 &  0.000 &  0.000 &  0.007 &  0.112 &  0.315 &  0.265 &  0.301 \\ 
Stroke & 8,298 &  0.001 &  0.004 &  0.011 &  0.092 &  0.269 &  0.256 &  0.368 \\ 
Fracture & 13,718 &  0.003 &  0.034 &  0.063 &  0.128 &  0.178 &  0.187 &  0.408 \\ 
Open wound & 3,642 &  0.002 &  0.047 &  0.097 &  0.197 &  0.224 &  0.167 &  0.266 \\ 
Appendicitis & 1,961 &  0.000 &  0.224 &  0.199 &  0.274 &  0.181 &  0.084 &  0.038 \\ 
Vehicle accident & 1,944 &  0.001 &  0.090 &  0.216 &  0.297 &  0.195 &  0.119 &  0.082 \\ 
Other accident & 9,782 &  0.003 &  0.033 &  0.034 &  0.082 &  0.154 &  0.208 &  0.486 \\ 

\bottomrule
\end{tabular}
\end{center}
\tabnote{\textwidth}{Notes: Table reports the number and age distribution of admissions, for different categories of admissions, over the time period March 13, 2020 through June 18, 2020. See \hyperref[app:cause]{Appendix \ref*{app:cause}} for definitions of the different causes of admissions (Cancer-other accident).}
\end{table}

\begin{table}
\caption{Test rates and prevalence bounds under monotonicity, by month and cause of admission}
\scriptsize
\label{tab:rates_bounds_group_narrow1}

\begin{center}
\begin{tabular}{l rrrr}\toprule
Month & \# Admits & \% Tested & Lower bound & Upper bound  \\
\midrule
\\
\multicolumn{5}{l}{\underline{A. Cancer}} \\
March & 585 &  0.079 &  0.002 &  0.022\\ 
April & 861 &  0.179 &  0.009 &  0.052\\ 
May & 986 &  0.385 &  0.005 &  0.013\\ 
June & 1,118 &  0.366 &  0.004 &  0.010\\ 
July & 1,154 &  0.377 &  0.004 &  0.011\\ 
August & 1,117 &  0.315 &  0.003 &  0.009\\ 
September & 1,146 &  0.325 &  0.001 &  0.003\\ 
October & 1,169 &  0.366 &  0.003 &  0.009\\ 
November & 1,065 &  0.441 &  0.022 &  0.049\\ 
December & 385 &  0.478 &  0.039 &  0.082\\ 

\\
\multicolumn{5}{l}{\underline{B. Labor and delivery}} \\
March & 908 &  0.007 &  0.000 &  0.000\\ 
April & 1,479 &  0.057 &  0.005 &  0.094\\ 
May & 1,519 &  0.203 &  0.009 &  0.042\\ 
June & 1,474 &  0.220 &  0.005 &  0.025\\ 
July & 1,650 &  0.247 &  0.003 &  0.012\\ 
August & 1,607 &  0.242 &  0.006 &  0.023\\ 
September & 1,452 &  0.174 &  0.003 &  0.016\\ 
October & 1,415 &  0.151 &  0.004 &  0.028\\ 
November & 1,334 &  0.331 &  0.030 &  0.090\\ 
December & 466 &  0.416 &  0.030 &  0.072\\ 

\\
\multicolumn{5}{l}{\underline{C. AMI}} \\
March & 456 &  0.184 &  0.022 &  0.119\\ 
April & 734 &  0.360 &  0.042 &  0.117\\ 
May & 877 &  0.437 &  0.042 &  0.097\\ 
June & 1,004 &  0.366 &  0.011 &  0.030\\ 
July & 1,005 &  0.423 &  0.013 &  0.031\\ 
August & 1,014 &  0.352 &  0.025 &  0.070\\ 
September & 1,046 &  0.334 &  0.015 &  0.046\\ 
October & 1,046 &  0.371 &  0.030 &  0.080\\ 
November & 1,062 &  0.459 &  0.078 &  0.170\\ 
December & 380 &  0.511 &  0.063 &  0.124\\ 

\\
\multicolumn{5}{l}{\underline{D. Stroke}} \\
March & 480 &  0.098 &  0.021 &  0.213\\ 
April & 772 &  0.206 &  0.019 &  0.094\\ 
May & 959 &  0.274 &  0.014 &  0.049\\ 
June & 938 &  0.247 &  0.010 &  0.039\\ 
July & 979 &  0.238 &  0.009 &  0.039\\ 
August & 986 &  0.208 &  0.007 &  0.034\\ 
September & 955 &  0.200 &  0.007 &  0.037\\ 
October & 1,017 &  0.280 &  0.015 &  0.053\\ 
November & 914 &  0.375 &  0.046 &  0.122\\ 
December & 298 &  0.399 &  0.037 &  0.092\\ 

\\
\multicolumn{5}{l}{\underline{D. Fractures and crushes}} \\
March & 628 &  0.048 &  0.003 &  0.067\\ 
April & 1,153 &  0.193 &  0.013 &  0.067\\ 
May & 1,517 &  0.365 &  0.019 &  0.052\\ 
June & 1,699 &  0.389 &  0.008 &  0.020\\ 
July & 1,706 &  0.371 &  0.008 &  0.021\\ 
August & 1,751 &  0.345 &  0.010 &  0.028\\ 
September & 1,722 &  0.398 &  0.009 &  0.022\\ 
October & 1,649 &  0.400 &  0.015 &  0.038\\ 
November & 1,428 &  0.472 &  0.038 &  0.080\\ 
December & 465 &  0.568 &  0.045 &  0.080\\ 

\bottomrule
\end{tabular}
\end{center}
\tabnote{\textwidth}{Notes: Table reports monthly number of admissions for cancer, along with the percent of admissions tested in hospital, and the bound on prevalence under test monotonicity. The lower bound is the confirmed positive rate and the upper bound is the positivity rate. See \hyperref[app:cause]{Appendix \ref*{app:cause}} for definitions.}
\end{table}

\begin{table}
\caption{Test rates and prevalence bounds under monotonicity, by month and cause of admission (continued)}
\scriptsize
\label{tab:rates_bounds_group_narrow2}
\begin{center}
\begin{tabular}{l rrrr}\toprule
Month & \# Admits & \% Tested & Lower bound & Upper bound  \\
\midrule
\\
\multicolumn{5}{l}{\underline{E. Open wounds}} \\
March & 180 &  0.106 &  0.011 &  0.105\\ 
April & 303 &  0.158 &  0.017 &  0.104\\ 
May & 411 &  0.282 &  0.024 &  0.086\\ 
June & 436 &  0.298 &  0.021 &  0.069\\ 
July & 480 &  0.283 &  0.008 &  0.029\\ 
August & 485 &  0.324 &  0.016 &  0.051\\ 
September & 424 &  0.377 &  0.026 &  0.069\\ 
October & 416 &  0.353 &  0.019 &  0.054\\ 
November & 385 &  0.384 &  0.044 &  0.115\\ 
December & 122 &  0.557 &  0.107 &  0.191\\ 

\\
\multicolumn{5}{l}{\underline{F. Vehicle accidents}} \\
March & 60 &  0.033 &  0.000 &  0.000\\ 
April & 134 &  0.134 &  0.000 &  0.000\\ 
May & 213 &  0.202 &  0.005 &  0.023\\ 
June & 278 &  0.266 &  0.000 &  0.000\\ 
July & 246 &  0.256 &  0.008 &  0.032\\ 
August & 291 &  0.265 &  0.010 &  0.039\\ 
September & 266 &  0.327 &  0.000 &  0.000\\ 
October & 219 &  0.306 &  0.009 &  0.030\\ 
November & 191 &  0.387 &  0.021 &  0.054\\ 
December & 46 &  0.478 &  0.022 &  0.045\\ 

\\
\multicolumn{5}{l}{\underline{G. Other accidents}} \\
March & 532 &  0.098 &  0.026 &  0.269\\ 
April & 844 &  0.210 &  0.033 &  0.158\\ 
May & 1,113 &  0.335 &  0.017 &  0.051\\ 
June & 1,150 &  0.310 &  0.012 &  0.039\\ 
July & 1,196 &  0.296 &  0.013 &  0.042\\ 
August & 1,215 &  0.288 &  0.012 &  0.040\\ 
September & 1,142 &  0.344 &  0.018 &  0.051\\ 
October & 1,166 &  0.346 &  0.031 &  0.089\\ 
November & 1,102 &  0.399 &  0.071 &  0.177\\ 
December & 322 &  0.506 &  0.096 &  0.190\\ 

\\
\multicolumn{5}{l}{\underline{H. Appendicitis}} \\
March & 122 &  0.033 &  0.016 &  0.500\\ 
April & 157 &  0.191 &  0.000 &  0.000\\ 
May & 229 &  0.376 &  0.022 &  0.058\\ 
June & 223 &  0.413 &  0.009 &  0.022\\ 
July & 225 &  0.409 &  0.009 &  0.022\\ 
August & 263 &  0.430 &  0.011 &  0.027\\ 
September & 248 &  0.399 &  0.016 &  0.040\\ 
October & 218 &  0.450 &  0.014 &  0.031\\ 
November & 194 &  0.485 &  0.046 &  0.096\\ 
December & 82 &  0.549 &  0.049 &  0.089\\ 

\bottomrule
\end{tabular}
\end{center}
\tabnote{\textwidth}{Notes: Table reports monthly number of admissions for cancer, along with the percent of admissions tested in hospital, and the bound on prevalence under test monotonicity. The lower bound is the confirmed positive rate and the upper bound is the positivity rate. See \hyperref[app:cause]{Appendix \ref*{app:cause}} for definitions.}
\end{table}

\clearpage 
	\section{Defining causes of admissions \label{app:cause}}

This section provides more details on our definition of ICLI, non-ICIL, and ``clear cause" hospitalization, listing the ICD-10 codes used to define each. 

Following \cite{afhsc2015}, the codes for influenza-like illness are 
B97.89, H66.9, H66.90, H66.91 H66.92, H66.93, J00, J01.9, J01.90, J06.9, J09, J09.X, J09.X1, J09.2, J09.X3, J09.X9, J10, J10.0, J10.00, J10.01, J10.08, J10.1, J10.2, J10.8, J10.81, J10.82, J10.83, J10.89, J11, J11.0, J11.00, J11.08, J11.1, J11.2, J11.8, J11.81, J11.82, J11.83, J11.89, J12.89, J12.9, J18, J18.1, J18.8, J18.9, J20.9, J40, R05, and R50.9. We say a hospitalizaiton is for an influenza-like illness if it has any of these diagnosis codes in any position. We say a hospitalization is for a COVID-like illness if it has any ICD-10 code among those that is among the CDC's lists of diagnosis codes for COVID-19 \cite{cdcCovidCodes}. These codes are J12.89, J20.8, J22, J40, J80, J98.8, O95.5, R05, R06.02, R50.9, U07.1, Z03.818, Z11.58, and Z20.828.

We define ICLI-related hospitalizations as ones with at least one ILI or CLI diagnosis code. We define non-ICLI related hospitalizations as hospitalized with diagnosis codes, but no ILI or CLI code.

We also define ``clear cause" hospitalizations. These are hospitalizations for labor and delivery, AMI, stroke, fractures and crushes, wounds, vehicle accidents, other accidents, appendicitis, or cancer. With the exception of cancer, we define a hospitalization as belonging to one of these groups if it has any diagnosis codes for that group, listed below. Cancer is treated differently because it can be a comorbidity. We say a hospitalization is for cancer if a cancer diagnosis (listed below) is an admitting diagnosis, the primary final diagnosis, or if chemotherapy diagnosis is present. We use the following ICD-10 codes.

\begin{itemize}
\item \textbf{AMI} I21, I22.
\item \textbf{Appendicitis} K35-K38.
\item \textbf{Cancer} C00-C97 (in primary or admitting diagnosis), or Z51.0-Z51.2 (in any position). 
\item  \textbf{Fracture/Crush} S02, S12, S22, S32, S42, S52, S62, S72, S82, S92, T02, S07, S17, S37, S47, S57,S67, S77, S87, S97, T07.
\item  \textbf{Labor and delivery} O60-O75, O80-O84.
\item  \textbf{Other accidents} W00-W99, X00-X59.
\item  \textbf{Stroke} I61-I64.
\item  \textbf{Vehicle accident} V01-V99.
\item  \textbf{Wound} S01, S11, S21, S31, S41, S51, S61, S71, S81, S91, T01.
\end{itemize}

\clearpage 
	\section{Calculating negative predictive  values with test-retest data \label{app:npv}}
	
\textbf{Setup and identification}
Here we show how to use data on multiple tests to simultaneously identify prevalence and test error rates, and how to use this information to obtain the negative predictive value (NPV) of at test under a narrow set of assumptions. Assume in particular that people have been tested exactly twice, with $R1_i$ the outcome of the first test and $R2_i$ the outcome of the second test for person $i$. Let $C_i$ be person $i$'s true infection status, which we assume is fixed between the tests.   Let $p = Pr(C_i = 1)$ be the prevalence of active SARS-CoV-2 infections in this twice-tested population. 

Test outcomes may differ from true infection status because of test errors. In general, therefore, there are four possible sequences of test outcomes: $(0,0), (0,1), (1,0), (1,1)$. We let $P_{ab} = Pr(R1_i = a, R2_i = b)$ for $(a,b) \in \left\lbrace 0, 1\right\rbrace^2$.

We make three strong assumptions to simplify the analysis. 

\begin{assumption} 
\label{spec_assumption}
The specificity of the test is 1. That is, $\beta = Pr(Rj_i=0|C_i=0)=1$. 
\end{assumption}

\begin{assumption} 
\label{sens_assumption}
The sensitivity of the test, $\alpha=Pr(Rj_i=1|C_i=1)$, does not depend on the initial test result.
\end{assumption}

\begin{assumption} 
\label{retest_assumption}
Retesting is random, i.e. independent of $R1_i$ and $C_i$.
\end{assumption}

Assumption \ref{spec_assumption} is the weakest of these assumptions. It implies that there are no false positives, which is consistent with typical practice \citep{ucsfDiagnostic}.  The remaining assumptions are stronger. Assumption \ref{sens_assumption} says that the test errors are independent of the initial test result. It would be violated, for example, if false negatives are more common for patients with high levels of mucus, and mucus levels are correlated across test results. Assumption \ref{retest_assumption} says that retesting rates do not depend on possible testing errors. We would expect this condition to fail if highly symptomatic people with negative tests are especially likely to test negative. We view this assumption as the most suspect.

Under these assumptions, the test outcome probabilities $P_{ab}$ simplify considerably. Since the probabilities sum to one,  and the assumptions imply that  $P_{10}=P_{01}$,  the only non-redundant probabilities are: 

\begin{align*}
P_{00} & = (1-p) + p(1-\alpha)^2 \\
P_{11} & = p \alpha^2.
\end{align*}

We can observe $P_{00}$ and $P_{11}$. Solving for the unknowns $p$ and $\alpha$, we have

\begin{align*}
p  & = \frac{ (P_{00}-P_{11} -1)^2  }{4 P_{11}  } \\
\alpha & = \frac{2 P_{11}}{1 - P_{00} + P_{11} }
\end{align*}

This shows how to get $p$ and $\alpha$ from two tests, and the assumption that specificity ($\beta$) equals 1. Our goal is to find the negative predictive value (NPV), which can be computed given knowledge of $\alpha, \beta$ and $p$. In general, for a single test $NPV = Pr(C_i=0|R_i=0)$. Applying Bayes rule shows that:

\begin{equation*}
NPV = \frac{1-p}{p(1-\alpha) + (1-p)}
\end{equation*}

\textbf{Results} 
To implement this approach, we construct a sample of all people who are tested on a given day, not tested the previous day, and then tested again in the next day. There are 835,195 such test pairs. We find $P_{00} = 0.884$ and $P_{11}= 0.113$. Nearly all the mass is on the diagonals; test results switch less than 1\% of the time. This fact, together with the assumption that specificity is equal to 1, implies very low false negative rates. Plugging these values into our formula, we have $p=0.116$ and $\alpha=0.987$, which implies $NPV = 0.998$.  Using instead, all people who are retested once within a three day period, we find similar results: $p=0.118, \alpha=0.972, NPV=0.996$.

We emphasize that these estimates are valid for the twice-tested population and under assumptions \ref{spec_assumption}-\ref{retest_assumption}, in particular, random retesting. The prevalence estimate is the prevalence among people tested twice, not the population prevalence. And it is only a valid estimate under assumptions 1-3. In reality, it is likely that retests are most common among suspected false negatives (i.e. when a highly symptomatic patient tests negative). We see some evidence for this: $P_{01}=0.0013$ and $P_{10}=0.0016$, a slight but significant difference implying that negative-then-positive is slightly more common than positive-than-negative, inconsistent with the random retesting assumption. We therefore do not view our estimates of prevalence and sensitivity as definitive; rather we think of the sensitivity estimate as a lower bound on sensitivity, because we have selected a retest sample which has a disproportionate number of false negatives. As $NPV$ is increasing in sensitivity, $\alpha$, our implied estimate of $1-NPV$ is likely an upper bound on $1-NPV$.

\clearpage
    \section{Confidence Intervals on Identification Regions \label{app:ci}}
    The main goal of this paper is to show how the combination of SARS-CoV-2 testing data and hospital records can be used to resolve some of the ambiguity in the prevalence of SARS-CoV-2 in the population. The focus is on ambiguity created because only a small fraction of the population is tested and it is likely that SARS-CoV-2 is more prevalent among people who are tested than among people who are not tested. We derive and compute upper and lower bounds on prevalence under different assumptions. Each of these bounds provide an expression of uncertainty about the prevalence of active SARS-CoV-2 infections that arises because the US does not have an organized program of surveillance testing based on testing random samples of the population.

A separate concern is that there may also be statistical uncertainty in the upper and lower bounds we report in the paper. To summarize the statistical uncertainty, we report 95\% confidence intervals around the identification region defined by the upper and lower bounds based on each assumption. To see the idea, let $\hat{L}$ and $\hat{U}$ be estimates of the population lower and upper bounds, $L$ and $U$. Now let $\hat{\sigma_L}$ and $\hat{\sigma_U}$ be estimates of the standard errors of the lower and upper bounds respectively. A 95\% confidence interval on the identification region is:

\begin{align}
\label{ci_formula}
[A(L), B(U)] &=  [\hat{L} - \hat{\sigma_{L}} \times 1.96 , \hat{U} + \hat{\sigma_{U}} \times 1.96]
\end{align}

The interval $[A(L), B(U)]$ is a 95\% confidence interval on the identification region in the sense that  $Pr(A(L) \leq L, U \leq B(U)) = .95 $ asymptotically.

To construct the confidence interval on each set of bounds in the paper, we estimate standard errors on each lower and upper bound. Estimating the standard errors is straightforward because each of the bounds can be re-expressed as a sample proportion and the conventional formula for the standard error of a proportion applies. 

\subsection{Standard Errors In The Covid Test Sample}

Under test monotonicity, the estimate of the lower bound on prevalence using the Indiana Covid Test data is simply the number of people who test positive expressed as a proportion of the state population:

\begin{align*}
\hat{L_{m}} & =  Pr(C_{it}|D_{it} = 1) Pr(D_{it} = 1) \\
& = \left( \frac{\sum_{i=1}^{N} T_{it}C_{it}}{\sum_{i=1}^{N}T_{it}} \right) \times \left(\frac{\sum_{i=1}^{N}T_{it}}{N}\right) \\
& = \frac{\sum_{i=1}^{N} T_{it}C_{it}}{N} \\
& = \frac{\text{Number of Positives}}{\text{Population of Indiana}}
\end{align*}

Likewise, the estimate of the upper bound under test monotonicity is simply the number of people who test positive expressed as a proportion of the number of people tested:

\begin{align*}
\hat{U_{m}} & =  Pr(C_{it}|D_{it} = 1) \\
&= \frac{\sum_{i=1}^{N} T_{it}C_{it}}{\sum_{i=1}^{N}T_{it}} \\
& = \frac{\text{Number of Positives}}{\text{Number Tested}}
\end{align*}

We apply the conventional approach to estimating the standard error of a sample proportion to obtain standard errors for the lower and upper bounds under test monotonicity. Letting $N$ represent the population of Indiana and $N_T$ represent the number of people tested,  we estimate the standard error of the lower and upper bounds using:

\begin{align*}
    \hat{\sigma_{L_{m}}} &= \sqrt{\frac{\hat{L_{m}}(1 - \hat{L_{m}})}{N}} \\
    \hat{\sigma_{U_{m}}} &= \sqrt{\frac{\hat{U_{m}}(1 - \hat{U_{m}})}{N_T}}
\end{align*}

These expression make it clear that the precision of our estimates of the upper and lower bounds depends on the size of the overall population being tested, the number of people who are tested in the population, as well as on the confirmed positive rate and positive testing rate.

We use the estimates of the standard errors of the upper and lower bounds to estimate a confidence interval using the expression in equation \ref{ci_formula}

\subsection{Standard Errors in the Hospital Sample}

We follow the same strategy to estimate standard errors on the upper and lower bounds in the hospital sample. Using $N_H$ to represent the number of people who are hospitalized for a given non-Covid condition, the lower and upper bounds in the hospital sample can also be expressed as proportions. The lower bound is:

\begin{align*}
\hat{L_{m}^{H}} & =  Pr(C_{it}|D_{it} = 1, H_{it} = 1) Pr(D_{it} = 1 | H_{it} = 1) \\
& = \frac{\sum_{i=1}^{N_H} T_{it}C_{it}}{N_H} \\
& = \frac{\text{Number of Positives in Hospital Sample}}{\text{Number in Hospital Sample}}
\end{align*}

And the upper bound is:

\begin{align*}
\hat{U_{m}^{H}} & =  Pr(C_{it}|D_{it} = 1, H_{it} = 1)  \\
& = \frac{\sum_{i=1}^{N_H} T_{it}C_{it}}{\sum_{i=1}^{N_H} T_{it}} \\
& = \frac{\text{Number of Positives in Hospital}}{\text{Number Tested in Hospital Sample}}
\end{align*}

The standard errors on these upper and lower bounds are:

\begin{align*}
    \hat{\sigma_{L_{m}^{H}}} &= \sqrt{\frac{\hat{L_{m}^{H}}(1 - \hat{L_{m}^{H}})}{N_H}} \\
    \hat{\sigma_{U_{m}^{H}}} &= \sqrt{\frac{\hat{U_{m}^{H}}(1 - \hat{U_{m}^{H}})}{N_{T}^{H}}}
\end{align*}

These expressions show that the standard error of the estimates of the upper and lower bounds in the hospital samples depend on the size of the hospitalized population, the number of people tested in the hospitalized population, as well as the confirmed positive rate, and positive testing rate in the hospitalized sample. Estimates of upper and lower bounds based on hospitalized sub-populations with smaller sample sizes will be less precisely estimated and will have larger standard errors and confidence intervals than estimates based on larger sub-populations.

With the standard errors in hand, we form confidence intervals around each set of test monotonicity bounds using equation \ref{ci_formula}.

\subsection{Age Adjustment}

Because the tested and hospitalized samples are not age representative of the general population, throughout the paper, we report both unadjusted results and age-standardized upper and lower bounds. This simply means that we stratify the data six age groups ( 0-17, 18-30, 30-50, 50-64, 65-74, and 75 and older) and then compute the upper and lower bounds within each age-strata. Afterwards, we average the age group specific bounds by weighting each age-specific bound by that age group's share of the Indiana population.

To estimate the standard error on the age standardized lower and upper bounds, we let $\hat{\sigma_{L_{a}}^{2}}$ and $\hat{\sigma_{U_{a}}^{2}}$ be the estimates of the asymptotic sampling variance of the lower and upper bounds in age group $a$. The standard error of the age standardized lower bound is $\hat{\sigma_L} = \sqrt{ \sum_{a=1}^{6} \left( \hat{\sigma_{L_{a}}^{2}} \times \pi(a) \right)} $, where $\pi(a)$ is the population share for age group $a$. Likewise, the standard error for the upper bound is $\hat{\sigma_U} = \sqrt{ \sum_{a=1}^{6} \left( \hat{\sigma_{U_{a}}^{2}} \times \pi(a) \right)} $

\end{document}